\documentclass{aa}
\usepackage{graphicx}

\newcommand{\aap}{A\&A}
\newcommand{\aaps}{A\&AS}
\newcommand{\apj}{ApJ}
\newcommand{\apjl}{ApJ}

\newcommand{\aj}{AJ}
\newcommand{\mnras}{MNRAS}

\newcommand{\pasj}{PASJ}

\newcommand{\lsol}{$L_{\odot}$}

\begin{document}
   \title{A Compact Array imaging survey of southern bright-rimmed
   clouds\thanks{Full versions of Figs.~1 and 2 may
   be found in the on-line supplement.}}

   \subtitle{}

   \author{M.A.~Thompson, J.S.~Urquhart \and G.J.~White}
   
   \offprints{M.A.~Thompson }

   \institute{Centre for Astrophysics \& Planetary Science,
              School of Physical Sciences,
              University of Kent,
              Canterbury,
              Kent CT2 7NR,
              UK\\
              \email{m.a.thompson@kent.ac.uk}
	       }

   \authorrunning{Thompson, Urquhart \& White}
   \titlerunning{A Compact Array imaging survey of BRCs}
   
   \date{}

   \abstract{We have carried out a radio-wavelength imaging survey of 45
  bright-rimmed clouds (BRCs), using the Australia Telescope Compact Array to
  characterise  the physical properties in their ionised boundary layers. We
  detected radio emission from a total of 25 clouds and using a combination of
  Digitised Sky Survey and mid-infrared MSX 8$\mu$m images classified the
  emission into that associated with the ionised cloud rims, that associated
  with embedded possible massive YSOs and that unlikely to be associated with
  the clouds at all. A total of 18 clouds display radio emission  clearly
  associated with the cloud rim and we determine the ionising photon flux
  illuminating these clouds and the electron density and pressure of their
  ionised boundary layers. Using a global estimate for the interior molecular
  pressure of these clouds we show that the majority are likely to be in
  pressure equilibrium and hence are currently being shocked by
  photoionisation-induced shocks. We identify those clouds where the predicted
  ionising photon flux is inconsistent with that derived from the observations
  and show that either the spectral types of the stars illuminating the BRCs
  are earlier than previously thought or that there must be additional ionising
  sources within the HII regions. Finally, we identify the radio sources
  embedded within the clouds with infrared stellar clusters and show that they
  contain late O and early B-type stars, demonstrating that a number of BRCs
  are intimately involved with high to intermediate-mass star formation.
  }

   \maketitle
%

\section{Introduction}

Bright rimmed clouds (BRCs) are found at the edges of HII regions and are the result
of the interaction of OB stars with their environment. The surface layers of BRCs are
photoionised and excited by nearby OB star(s), resulting in a bright optical rim
located on the side of the cloud facing the OB star. The cloud material is often
swept into a cometary morphology by the radiation pressure (e.g.~Bertoldi
\cite{bertoldi};  Lefloch \& Lazareff \cite{ll94}). Dense cores found at the head of
the clouds may  shield the remaining cloud material from the UV radiation, resulting
in a column or finger structure similar to that observed in the Eagle Nebula (Hester
et al.~\cite{hester96}; White et al.~\cite{whiteeagle}). These phenomena lead to the
other common names for bright-rimmed clouds such as elephant trunk nebulae, cometary
globules or speck globules (Bertoldi \cite{bertoldi}).

The photoionisation of the cloud surface has been linked to induced star formation within the clouds
(e.g.~Elmegreen \cite{e91}; Sugitani et al.~\cite{sfo}; Sugitani et al.~\cite{sfmo89}). In a process
known as Radiative-driven Implosion (or RDI) shocks driven into the cloud by the photoionisation of its
outer layers compress the molecular gas, forming dense cores at the head of the cloud and possibly
triggering their collapse. The physical processes behind RDI are relatively well-understood and have
been the subject of recent theoretical studies  (Bertoldi \cite{bertoldi}; Bertoldi \& McKee
\cite{bk90}; Lefloch \& Lazareff \cite{ll94}, \cite{ll95}). The radiative-driven implosion of molecular
clouds at the periphery of HII regions may amount to a cumulative total of some several hundred stars
per HII region (Ogura et al.~\cite{osp02}) and perhaps 15\% or more of the
low-to-intermediate mass stellar mass function (Sugitani et al.~\cite{sfo}).

One of the key diagnostic indicators in the RDI models is the ratio of external
to internal pressure. As the surface layers of the BRC become ionised a
recombination layer known as the \emph{ionised boundary layer} (or IBL) develops
on the side of the cloud facing the ionising star. If the conditions within the
IBL are such that the pressure within the ionised exterior gas is greater than
or equal to the interior neutral cloud pressure, photoionisation-induced shocks
and a D-critical ionisation front propagate into the cloud interior, compressing
and heating the molecular gas (Bertoldi \cite{bertoldi}; Lefloch \& Lazareff
\cite{ll94}). On the other hand, if the BRC is overpressured with respect to the
ionised boundary layer then the ionisation front stalls at the cloud surface
until the increasing recombination within the IBL raises the ionised gas
pressure to equilibrium with the interior cloud pressure (Lefloch \& Lazareff
\cite{ll94}). As equilibrium is reached the evolution of the cloud follows the
same path as the initially underpressured case.

The RDI models show that the evolution of a BRC critically depends upon the
\emph{duration} of its UV illumination.  The pressure balance between the interior
and exterior of the BRC thus allows us to discriminate between clouds that are
currently affected by photoionisation-induced shocks and those clouds that are as yet
relatively unaffected. The physical conditions in the ionised boundary layer are
crucial to understand the RDI process and any subsequent star formation, as it is the
ionised boundary layer which traces the interaction between the UV radiation from the
OB star(s) and the molecular gas of the bright-rimmed cloud. 
However, although at least 89 BRCs have been identified as potential star-forming
regions via IRAS searches (Sugitani et al.~\cite{sfo}; Sugitani \& Ogura
\cite{so94}) only a handful of clouds have had the conditions in their ionised
boundary layers  determined or their pressure balance investigated (Thompson et
al.~\cite{thompson03a}; White et al.~\cite{whiteeagle}; Megeath \& Wilson
\cite{mw97}; Lefloch et al.~\cite{llc97}).

Radio continuum observations are a valuable probe of the conditions within the
ionised boundary layer. From the radio free-free emission we can determine the
ionised gas pressure, the ionising photon flux impinging upon the cloud and the
electron density of the ionised boundary layer (Thompson et al.~\cite{thompson03a};
Lefloch et al.~\cite{llc97}). Data from the NRAO VLA Sky Survey (NVSS;
Condon et al.~\cite{nvss}) may be used for BRCs north of $\delta = -40$\degr\
(e.g.~Thompson et al.~\cite{thompson03a}) but no such resource\footnote{The Sydney
University Molonglo Sky Survey (SUMSS; Bock et al.~\cite{bls99}) has
comparable resolution and sensitivity to NVSS for $\delta \le -30$\degr\ but does not
yet cover a substantial portion of the southern galactic plane.} currently exists for
the BRCs in the southern hemisphere survey of Sugitani \& Ogura (\cite{so94}). As a
first step toward establishing the physical conditions in the ionised boundary layers
and pressure balances for a wide sample of BRCs we present an Australia Telescope
Compact Array\footnote{The Australia Telescope Compact Array is funded by the Commonwealth of
Australia for operation as a National Facility  managed by CSIRO.} (ATCA) radio imaging survey of the BRCs from the Sugitani \& Ogura
(\cite{so94}) catalogue.

The observations and data reduction procedure are described in
Sect.~\ref{sect:obsvns}. The data from the survey are presented and analysed in
Sect.~\ref{sect:analysis}, with particular reference to the likely nature of the
detected emission (i.e.~does the emission emanate from the ionised boundary
layer or from embedded sources?).  In Sect.~\ref{sect:discuss} we discuss the
implications of the survey and estimate the likely ratio between external and
internal pressure of the BRCs  detected in the survey, using assumptions for the
internal molecular gas pressure. Finally in Sect.~\ref{sect:concln} we present a
summary of our results and conclusions.


\section{Observations and data reduction}
\label{sect:obsvns}

We obtained radio images of the 45 BRCs from the southern hemisphere  survey of
Sugitani \& Ogura (\cite{so94})  using the Australia Telescope Compact Array
(ATCA), which is located
at the Paul Wild Observatory, Narrabri, New South Wales, Australia. The observations were carried out on the 9th
January, the 16th January and the 29th June 2002. The ATCA is a east-west array
comprising 6 $\times$
22 m antennas which may be positioned in several configurations along a railway
track. The sixth antenna is located three kilometres further west of the remaining
five antennas. Each antenna is fitted with a dual feedhorn system allowing
simultaneous dual-frequency observations at either 20/13 cm or 6/3 cm. 

\begin{table}[ht!]
\begin{center}
\caption{Clouds observed in the survey, the pointing positions used and the resulting
image r.m.s.~noise values. Each cloud is identified by its SIMBAD ID, which corresponds 
to
the cloud number in Sugitani \& Ogura (\cite{so94}) }
\label{tbl:sources}
\begin{tabular}{cccccc}\hline\hline
Cloud ID & \multicolumn{2}{c}{Primary beam centre} &  $\lambda$ & Image rms\\\
&$\alpha$(2000)&$\delta$(2000) &(cm) &  (mJy/beam)\\
\hline
SFO 45 & 07:18:23.7 & -22:06:13  & 20/13  & 1.72/0.73\\
SFO 46 & 07:19:25.8 & -44:35:01  & 20/13  & 1.45 /0.24\\
SFO 47 & 07:31:48.9 & -19:27:33  & 20/13  & 1.10/0.60\\
SFO 48 & 07:34:24.9 & -46:54:12  & 20/13  & 0.86/0.57 \\
SFO 49 & 07:35:41.0 & -18:48:59  & 20/13  &  1.42/1.62\\
SFO 50 & 07:40:28.5 & -43:06:20  & 20/13  & 1.2/0.50\\
SFO 51 & 08:09:32.8 & -36:05:00  & 20/13  & 0.94/0.28\\
SFO 52 & 08:25:43.9 & -51:00:37  & 20/13  & 3.15/1.05 \\
SFO 53 & 08:26:31.7 & -50:40:30  & 20/13  & 3.20/3.60 \\
SFO 54 & 08:35:31.7 & -40:38:28  & 20/13  & 1.70/0.53 \\
SFO 55 & 08:41:13.0 & -40:52:03  & 20/13  & 0.50/0.50\\
SFO 56 & 08:42:59.6 & -39:59:56  & 20/13  & 1.20/0.30\\
SFO 57 & 08:44:07.7 & -41:16:14  & 6/3    & 0.67/0.33\\
SFO 58 & 08:45:25.4 & -41:16:02  & 20/13  &0.61/0.33\\
SFO 59 & 08:58:04.2 & -47:22:57  & 20/13  & 83.40/101.58\\
SFO 60 & 09:00:01.2 & -47:31:37  & 20/13  & 120/200\\
SFO 61 & 11:00:13.5 & -59:36:09  & 6/3    & 0.50/0.18 \\
SFO 62 & 11:01:15.8 & -59:51:01  & 6/3    & 0.40/0.21 \\
SFO 63 & 11:03:18.8 & -59:48:01  & 6/3    & 0.68/0.34\\
SFO 64 & 11:12:18.1 & -58:46:20  & 20/13  & 1.00/0.52\\
SFO 65 & 11:32:55.3 & -63:27:59  & 20/13  & 3.10/1.30\\
SFO 66 & 11:33:49.5 & -63:16:20  & 6/3    & 0.67/0.32 \\
SFO 67 & 11:34:00.7 & -63:11:19  & 6/3    &0.67/0.32\\
SFO 68 & 11:35:31.9 & -63:14:51  & 6/3    &0.80/0.33\\
SFO 69 & 11:41:11.2 & -63:23:13  & 20/13  &4.98/5.00\\
SFO 70 & 11:42:11.2 & -63:07:50  & 6/3    &0.51/0.33\\
SFO 71 & 13:08:12.7 & -62:10:27  & 20/13  &7.10/4.10\\
SFO 72 & 13:19:07.7 & -62:33:44  & 20/13  &3.60/1.80\\
SFO 73 & 13:20:05.7 & -62:24:03  & 20/13  &1.00/6.02 \\
SFO 74 & 14:19:42.1 & -61:25:17  & 20/13  &1.67/0.6\\
SFO 75 & 15:55:50.4 & -54:38:58  & 20/13  &41.00/9.00\\
SFO 76 & 16:10:38.6 & -49:05:52  & 6/3    & 0.32/1.02\\
SFO 77 & 16:19:53.9 & -25:33:39  & 6/3    &0.67/0.22\\
SFO 78 & 16:20:52.9 & -25:08:07  & 6/3    &0.53/0.23\\
SFO 79 & 16:40:00.1 & -48:51:45  & 6/3    &43.33/41.02\\

SFO 80 & 16:40:16.8 & -48:42:25  & 20/13  &10.03/8.62\\
SFO 81 & 16:41:08.8 & -49:17:45  & 20/13  &6.59/2.30\\
SFO 82 & 16:46:50.2 & -41:13:54  & 20/13  & 0.50/0.42\\
SFO 83 & 16:52:12.0 & -40:48:09  & 6/3    & 0.51/0.27\\
SFO 84 & 16:53:45.1 & -40:07:22  & 20/13  & 3.99/1.00\\
SFO 85 & 16:59:06.4 & -42:42:04  & 20/13  & 13.00/67.00\\
SFO 86 & 17:49:36.1 & -31:29:13  & 20/13  & 3.52/1.00 \\
SFO 87 & 18:02:51.6 & -24:22:08  & 20/13  & 42.01/57.11 \\
SFO 88 & 18:04:16.8 & -24:06:59  & 20/13  & 26.50/12.1\\
SFO 89 & 18:09:54.3 & -24:04:56  & 20/13  & 5.12/1.54\\
\hline 
\end{tabular}
\end{center}
\end{table}

\begin{table}[ht!]
\begin{center}
\caption{Observing parameters for the ATCA radio imaging survey. All observing dates were
 in 2002.}
\label{tbl:obspar}
\begin{tabular}{llll}\hline\hline
Observing dates & 9th Jan & 16th Jan & 29th Jun \\ \hline
Configuration & 750A & 750A & EW214 \\
Wavelength & 20/13 cm & 20/13 cm & 6/3 cm \\
Bandwidth & 128 MHz & 128 MHz & 128 MHz \\
Time per source & 30 min & 30 min & 30 min \\
Phase calibrators & 1109$-$56 & 0647$-$75& 0906$-$47\\
 & 1352$-$63 & 0733$-$174& 1109$-$56\\
 & 1613$-$586 & 0826$-$373& 1622$-$297\\
 & 1646$-$50 & 0906$-$47& 1646$-$50\\
 & 1748$-$253 & & \\ \hline
\end{tabular}
\end{center}
\end{table}

One of our main considerations in planning the survey observations  was to maximise
the largest angular scale present in the final images.  Interferometers act as a high
pass filter removing Fourier components that are less than the minimum antenna
spacing. In order to maximise the largest angular scale present in the image one
should ideally observe with the most compact array possible, at the longest possible
wavelength and over a full track in the $uv$ plane. Practical considerations usually
enforce a compromise in one or more of these conditions. The optical diameters of
the bright-rimmed clouds are typically 4--5\arcmin\ as measured from POSS-II optical
plates by Sugitani \& Fukui (\cite{so94}) and so we required a largest angular scale
of at least this size. 

At the start of the survey the most compact suitable configuration was the 750 m array,
which for a series of cuts through the $uv$-plane   yields a largest angular scale of
$\sim$ 3\farcm6 at a wavelength of 13 cm and $\sim$ 5\farcm6 at a wavelength of 20 cm.
Thus, in order to maintain the necessary largest scale we begun the survey using the
simultaneous 20/13 cm mode of the ATCA. The shorter 6/3 cm mode is preferable to avoid
significant contamination by steep-spectrum background sources, but even with the more
compact 367 m array the resulting largest angular scale is smaller than the typical BRC
size (as estimated from the Digitised Sky Survey images). A total of 32 clouds were
observed at 20/13 cm using the 750 m array configuration. However, mid-way through the
radio imaging survey a new, more compact 214 m array was successfully trialed and the
remaining 13 clouds were observed at 6/3 cm using this configuration.

Each cloud from the Sugitani \& Ogura (\cite{so94}) survey was observed with a single
pointing with the primary beam centre pointed toward the IRAS point source position.
Table \ref{tbl:sources} shows the clouds, their positions and the wavelengths  that
were observed. To maximise the $uv$ coverage the integration time for  each cloud
was  split into a series of 6 $\times$ 5 minute ``cuts'' which were spread over a wide
range of hour angles. The resulting theoretical
r.m.s.~sensitivities for this integration time are 0.15, 0.2, 0.15 and 0.2 mJy/beam
for 20, 13, 6 and 3 cm respectively (assuming natural weighting). Sidelobes from
strong confusing sources in the primary beam may raise this considerably as can be seen in
Table \ref{tbl:sources}. We  will elaborate upon this issue further in 
in Sect.~\ref{sect:analysis}. The full receiver bandwidth of 128 MHz was used for
both frequency pairs, split into 32 channels. Phase calibrators were observed for a
few minutes approximately once every hour and the primary ATCA flux calibrator
1934$-$638 was observed at least  once per observing session. The observing
parameters of the survey are summarised in Table \ref{tbl:obspar}

The data were reduced and calibrated using the synthesis reduction package \emph{Miriad} 
(Sault et al.~\cite{miriad}) and following standard ATCA procedures.  The
images were CLEANed using a robust weighting of 0 to simultaneously minimise the r.m.s.~noise in
the images and maximise their sensitivity to large angular scales. The long (in excess of 3 km)
baselines  from the 6th antenna were found to significantly worsen the resulting images
and data from this antenna were excluded from the final cleaned images to improve the
large-scale sensitivity. The resulting synthesised beamwidths are typically $\sim$
90\arcsec\ for the 20 and 6 cm images and $\sim$ 60\arcsec\ for the 13 and 3 cm images.
The synthesised beamwidths are a strong function of the  declination -- images taken
close to the  celestial equator possess  elliptical synthesised beams, in some cases
with a semi-major FWHM  of up to $\sim$ 150\arcsec\ for images at a wavelength of 20 cm.

\section{Results and analysis}
\label{sect:analysis}

\subsection{Source identification and classification} 
\label{sect:sourceid}

\renewcommand{\thefootnote}{\alph{footnote}}
\setcounter{footnote}{0}
\begin{table*}
\begin{center}
\caption{Identifiers, positions, peak flux and integrated flux densities for BRC sources
detected at 3 or 6 cm. 3$\sigma$ upper limits for non-detections are indicated. }
\label{tbl:6/3sources}
\begin{minipage}{\linewidth}
\begin{tabular}{lccccccl}\hline\hline
	&  &  		 	 & \multicolumn{2}{c}{Peak flux} (mJy/beam)   &
	\multicolumn{2}{c}{Integrated flux} (mJy)   \\
Source ID	&  $\alpha$(2000)	 & 	$\delta$(2000)		& 3 cm 	& 6 cm &
3 cm & 6 cm & Type\\
\hline
\object{SFO 57} &    08:44:11.4 & -41:16:17 & 	8.0	&	30.7 & 33.4 & 76.4  & 1\\
\object{SFO 61a} &   11:00:08.0 & -59:36:06 & 1.9    &    7.1\footnotemark
\setcounter{footnote}{0}   & 2.1  &  16.2\footnotemark \setcounter{footnote}{1} & 1\\
\object{SFO 61b} &   11:00:08.0 & -59:37:48 & 1.5  & \ldots		 & 1.7 &
\ldots & 4\\		 
\object{SFO 62} &    11:01:16.6 & -59:50:40&  	   68.2   &	   126  & 112 & 197 & 2\\
\object{SFO 63} &    11:03:24.8 & -59:48:55&  	   11.5   &	   33.6   & 19.0 & 72.9 & 1
 \\
\object{SFO 66a}&    11:33:51.2 & -63:15:56&  	 3.4	 &	 4.3\footnotemark
\setcounter{footnote}{1}   & 3.4 & 4.3\footnotemark \setcounter{footnote}{1} & 1 \\
\object{SFO 66b}&    11:33:36.1 & -63:16:32&  	   1.7\footnotemark
\setcounter{footnote}{1}    &
    9.6   & 1.7\footnotemark \setcounter{footnote}{0} & 9.6\footnotemark
    \setcounter{footnote}{1} & 4\\
\object{SFO 66c}&    11:33:15.6 & -63:17:26&  	  54.7   &	  111  & 56.3& 105 & 4\\
\object{SFO 66d}&    11:33:07.3 & -63:13:14&  2.4    &       $<$
1.1\footnotemark \setcounter{footnote}{1}  & 3.4 & $<$ 1.1\footnotemark
\setcounter{footnote}{1} & 4\\
\object{SFO 67a}&    11:34:04.1 & -63:12:11&  	   $<$ 0.9  &	   16.3   & $<$ 0.9 & 52.4 & 1 \\
\object{SFO 67b}&    11:34:07.3 & -63:18:56&  	  $<$ 0.9  &	  10.9   & $<$ 0.9 & 8.0 & 4 \\
\object{SFO 68} &    11:35:32.4 & -63:15:33&  	   16.6   &	   45.6   & 32.4  & 149 & 1 \\
\object{SFO 70} &    11:42:04.1 & -63:08:11&  	  2.3    &	  4.8   & 7.0 & 5.6  & 1  \\
\object{SFO 76} &    16:10:37.6 & -49:05:46&  	  62.5   &	  141.5  & 130.4 & 400.0 & 1 \\
\object{SFO 77} &    16:19:56.5 & -25:33:51&  	 15.1   &	 42.7   & 40.5 & 91.5 & 1 \\
\object{SFO 78} &   16:20:44.9 & -25:04:40&  2.3    &       $<$ 0.7   & 12.7 & $<$ 0.7 & 4 \\
\object{SFO 79} &    16:39:59.8 & -48:51:42&  	 4100	 &	 6000\footnotemark
\setcounter{footnote}{1}   & 4800  & 6000\footnotemark \setcounter{footnote}{1}  & 3 \\
\object{SFO 83} &    16:52:13.5 & -40:48:30&  	   4.5    &	   14.2   & 6.6  & 23.4 & 1 \\       
\hline
\end{tabular}
\footnotetext[1]{SFO 61a and SFO 61b cannot be separated at 20cm. The quoted
fluxes are for the combined source observed at 20cm, which has a peak position
of $\alpha_{2000}$ = 11$^{\rm h}$00$^{\rm m}$08\fs8, $\delta_{2000}$ =
$-$59\degr30\arcmin48\arcsec.}
\footnotetext[2]{Source is unresolved at this wavelength}
\end{minipage}
\end{center}
\end{table*}

\renewcommand{\thefootnote}{\alph{footnote}}
\setcounter{footnote}{0}
\begin{table*}
\begin{center}
\caption{Identifiers, positions, peak flux and integrated flux densities for BRC sources
detected at 13 or 20 cm. 3$\sigma$ upper limits for non-detections are indicated.}
\label{tbl:20/13sources}
\begin{minipage}{\linewidth}
\begin{tabular}{lccccccl}\hline\hline
	&  &  		 	 & \multicolumn{2}{c}{Peak flux} (mJy/beam)   &
	\multicolumn{2}{c}{Integrated flux} (mJy)   \\ 
Source ID	&  $\alpha$(2000)	 & 	$\delta$(2000)		& 13 cm 	&
20 cm & 13 cm & 20 cm & Type\\\hline 
\object{SFO 47a} & 07:32:10.0 & -19:28:48&	11.7 &	      14.4      & 15.0 & 24.9 & 4\\
\object{SFO 47b} & 07:31:49.0 & -19:24:38&	28.8 &		13.5	& 93.8& 186 & 4\\
\object{SFO 49a} &  07:36:02.5 & -18:53:16 &  242& 	316	&274 & 357 & 4	\\
\object{SFO 49b} & 07:35:34.5 & -18:45:32 &  194& 	294.2	& 463 & 568  & 4 \\
\object{SFO 56} &  08:42:57.0 & -40:00:26 & $<$ 3.6	&
5.2\footnotemark\setcounter{footnote}{0}	& $<$ 3.6 &
5.2\footnotemark\setcounter{footnote}{0} & 4 \\
\object{SFO 58a} &  08:45:15.8 & -41:13:33 &		 2.7  &	$<$ 1.0 	& 2.7 & $<$ 1.0 & 4 \\
\object{SFO 58b} &  08:45:19.5 & -41:15:13 &		 6.9  &	$<$ 1.0 	& 8.7 & $<$ 1.0 &  1 \\
\object{SFO 59} &  08:58:9.1& -47:21:51 &      3100	&	$<$ 305	& 10400 & $<$ 305 & 3 \\ 
\object{SFO 64} &  11:12:13.7 & -58:46:45& 	14.4	&	45.6	& 117 & 236 & 1\\
\object{SFO 69} &  11:41:04.9 & -63:22:05& 	25.4	&	$<$ 15.0	& 133 & $<$ 15.0 & 4\\
\object{SFO 74}&   14:19:40.9 & -61:25:17&	9.9	&	25.0	& 26.3 & 37.3  & 3\\
\object{SFO 75}&   15:55:43.4 & -54:39:07&	115.0	&	443	& 247 & 479 & 1
\\
\object{SFO 82a}&  16:47:22.1 & -41:13:06& 	3.0\footnotemark\setcounter{footnote}{0}	&	12.3	& 3.0\footnotemark\setcounter{footnote}{0}& 36.7 & 1 \\
\object{SFO 82b}&  16:47:19.4 & -41:15:43&	5.1	&	23.2	& 6.7 & 63.6 & 1	\\
\object{SFO 84} &  	16:53:51.9 & -40:08:09&	
\ldots \setcounter{footnote}{1}\footnotemark\setcounter{footnote}{1}	&	11.7	&
\ldots \footnotemark\setcounter{footnote}{1} & 28.6 & 1 \\
\object{SFO 85a} & 16:59:04.0&  -42:41:46& 40.3\footnotemark\setcounter{footnote}{0} & $<$ 201 & 40.3\footnotemark\setcounter{footnote}{0} & $<$ 201 &  3\\
\object{SFO 85b} & 16:59:12.0&  -42:43:15& 67.4 & $<$ 201 & 97.2 & $<$ 201 &  4\\
\object{SFO 89}&         18:09:48.9 & -24:04:56 & 		63.2	&	102	& 28.6 &
369 & 1\\
\hline
\end{tabular}
\footnotetext[1]{Source is unresolved at this wavelength}
\footnotetext[2]{In the cleaned 13 cm image SFO 84 lies in an area affected by a 
striping artefact caused by the cleaning process. It is not possible to set a
sensible flux limit for this source at 13 cm.}
\end{minipage}
\end{center}
\end{table*}

We have mapped the radio emission around 45 bright-rimmed clouds from the survey of
Sugitani \& Ogura (\cite{so94}). For the purposes of our survey we focus solely upon the
radio emission that is associated with each bright-rimmed cloud, rather than the many
unassociated and potentially background point sources that are predominantly found in the
20 cm images. We identified radio
sources in each of the final cleaned images as features with peak fluxes greater than 3
times the r.m.s.~noise in the image (i.e.~$\ge 3\sigma$) and positionally associated with
the optical clouds seen in the Digitised Sky Survey images of Sugitani \& Ogura
(\cite{so94}).

We detected no radio emission to a level of 3 times the r.m.s.~noise associated
with the rims of the clouds SFO 45, 46, 48, 50--55, 60, 71--73, 80, 81, 83, or
86--88. We calculate upper limits for the ionising flux illuminating these
clouds in Sect.~\ref{sect:uplim}.  A total of 36 radio sources positionally
associated with the bright-rimmed clouds were detected. The coordinates,  peak
fluxes and integrated flux densities of the 36 identified radio sources are
found in Tables \ref{tbl:6/3sources} and \ref{tbl:20/13sources}. The peak and
integrated flux densities were measured from the final cleaned images using the
visualisation package \emph{kview} (Gooch et al.~\cite{kview})

Given the large density of extragalactic radio sources on the sky at 20 cm, it is 
important to consider the likelihood that the emission we detect positionally coincident
with the clouds may originate from background extragalactic sources. Using the source counts from the
NVSS (Condon et al.~\cite{nvss}), which has a similar synthesised beamwidth and
sensitivity limit to our observations, and assuming that the source counts are similar for both
hemispheres, we estimate that no more than 4 BRC-associated sources in our survey are 
likely to be background extragalactic sources. 

We attempted to rule out background sources by measuring the spectral index of the emission, however
the ATCA is not a scaled array between the two wavelengths that were observed simultaneously and the
measured spectral index is thus highly dependent upon the $uv$ coverage of the observations and the
quality of the ``dirty beam'' (Cornwell et al.~\cite{cbb99}). Even a small uncertainty in the source
flux at each wavelength due to the nonlinear CLEAN algorithm can result in a large uncertainty in the
spectral index and we were unable to use the measured spectral indices to discriminate between the
flat-spectrum emission expected for thermal free-free radiation or the steep non-thermal spectra
expected from background radio galaxies.

In order to aid the classification of the detected radio sources and rule out
any likely chance associations with background extragalactic sources we compared
the radio emission to optical DSS images and mid-infrared Midcourse Space
Experiment (MSX) 8 $\mu$m Band A images.  The MSX satellite performed a
mid-infrared survey of the Galactic plane at a spatial resolution of $\sim$
18\arcsec\, with simultaneous observations in  four wavelength bands between 6
and 25 $\mu$m (Price et al.~\cite{price01}). Band A of the MSX survey spans a
wavelength range of 6.8--10.8 $\mu$m, which includes emission bands attributed
to  Polycyclic Aromatic Hydrocarbons (PAHs) at 7.7 and 8.6 $\mu$m. PAH emission
is known to be a good tracer of UV-irradiated photon-dominated regions (Leger \&
Puget \cite{leger84}) as the PAHs are transiently heated by the absorption of UV
photons. Comparing the appearance of the clouds in the radio, optical and
mid-infrared allows the morphology of the ionised rim and PDR in the underlying
molecular cloud to be traced irrespective of visual extinction or contamination
effects; discriminating between chance associations and true bright-rim
emission.  

\begin{figure*}[ht!]
\includegraphics*[scale=0.63,trim=0 0 30 0]{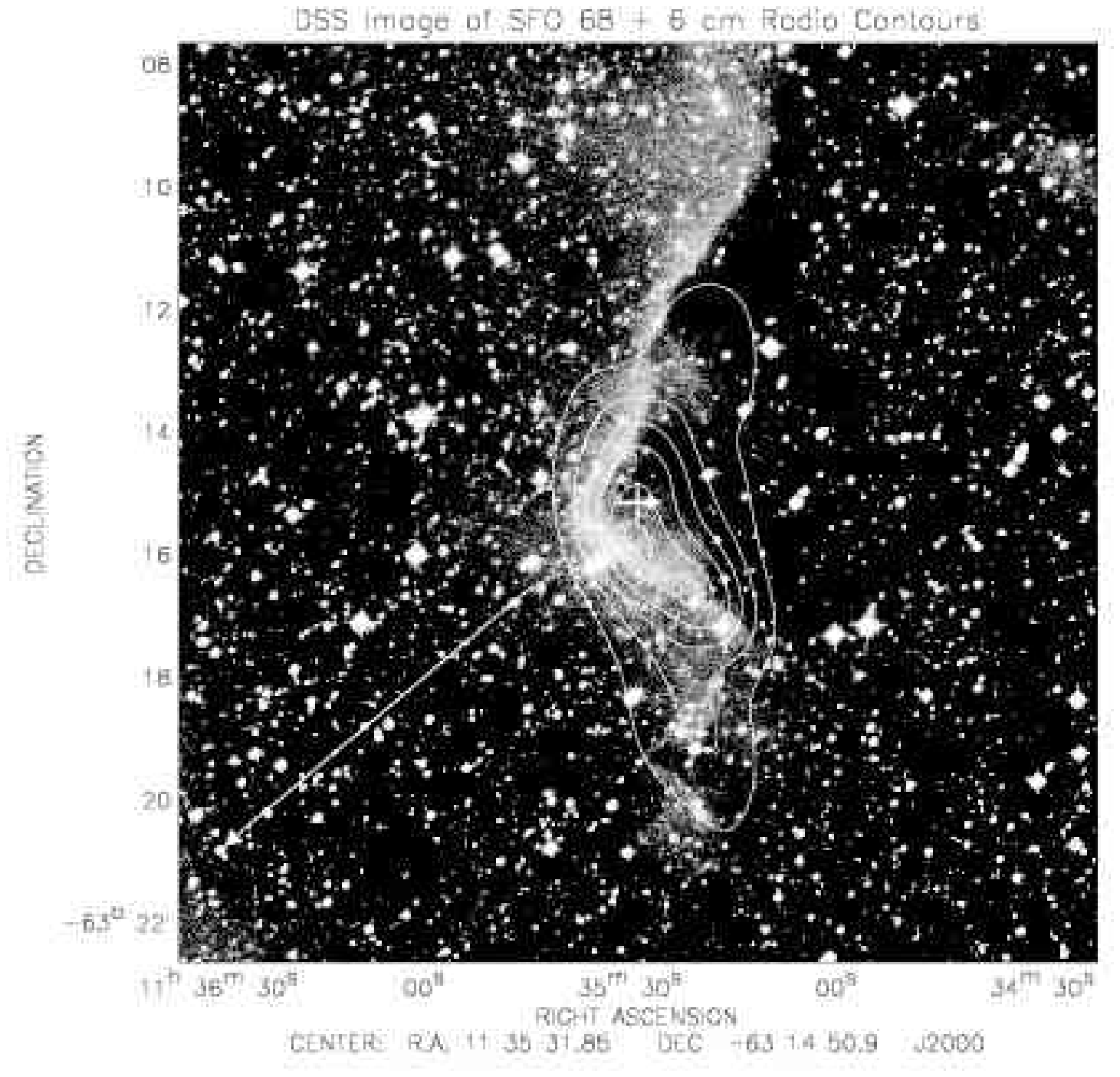}
\includegraphics*[scale=0.63,trim=0 0 30 0]{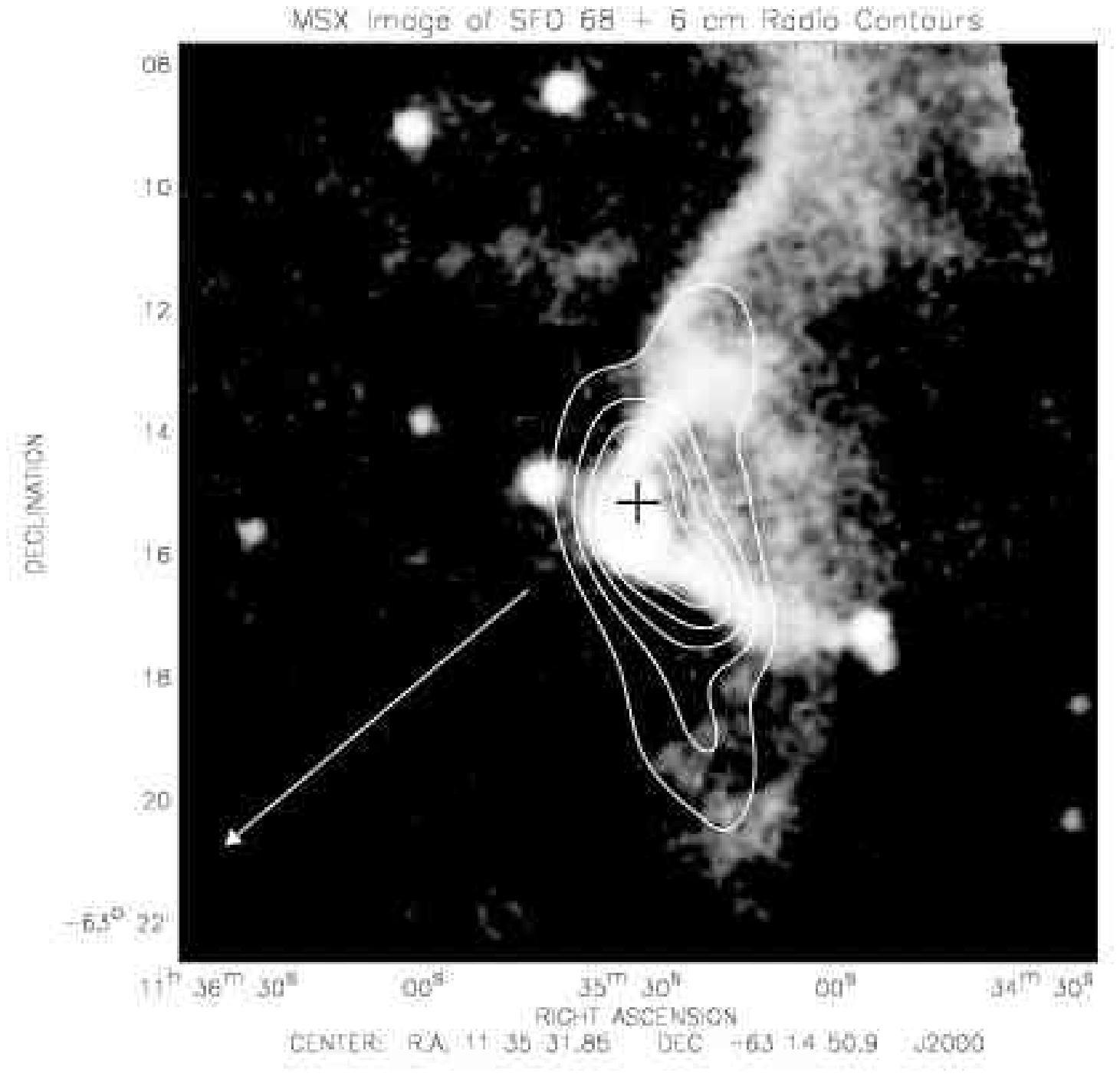}\\
\includegraphics*[scale=0.63,trim=0 0 30 0]{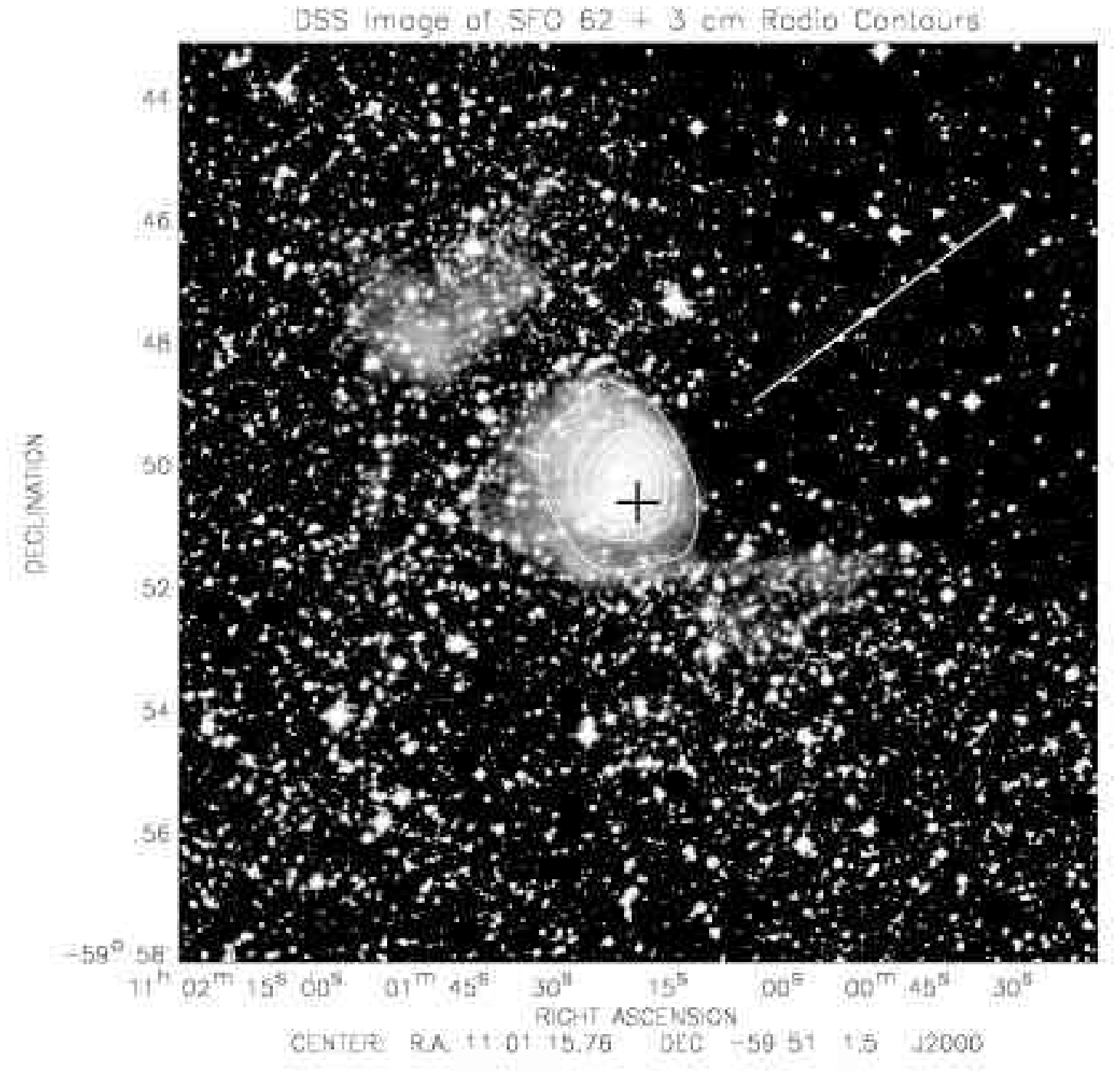}
\includegraphics*[scale=0.63,trim=0 0 30 0]{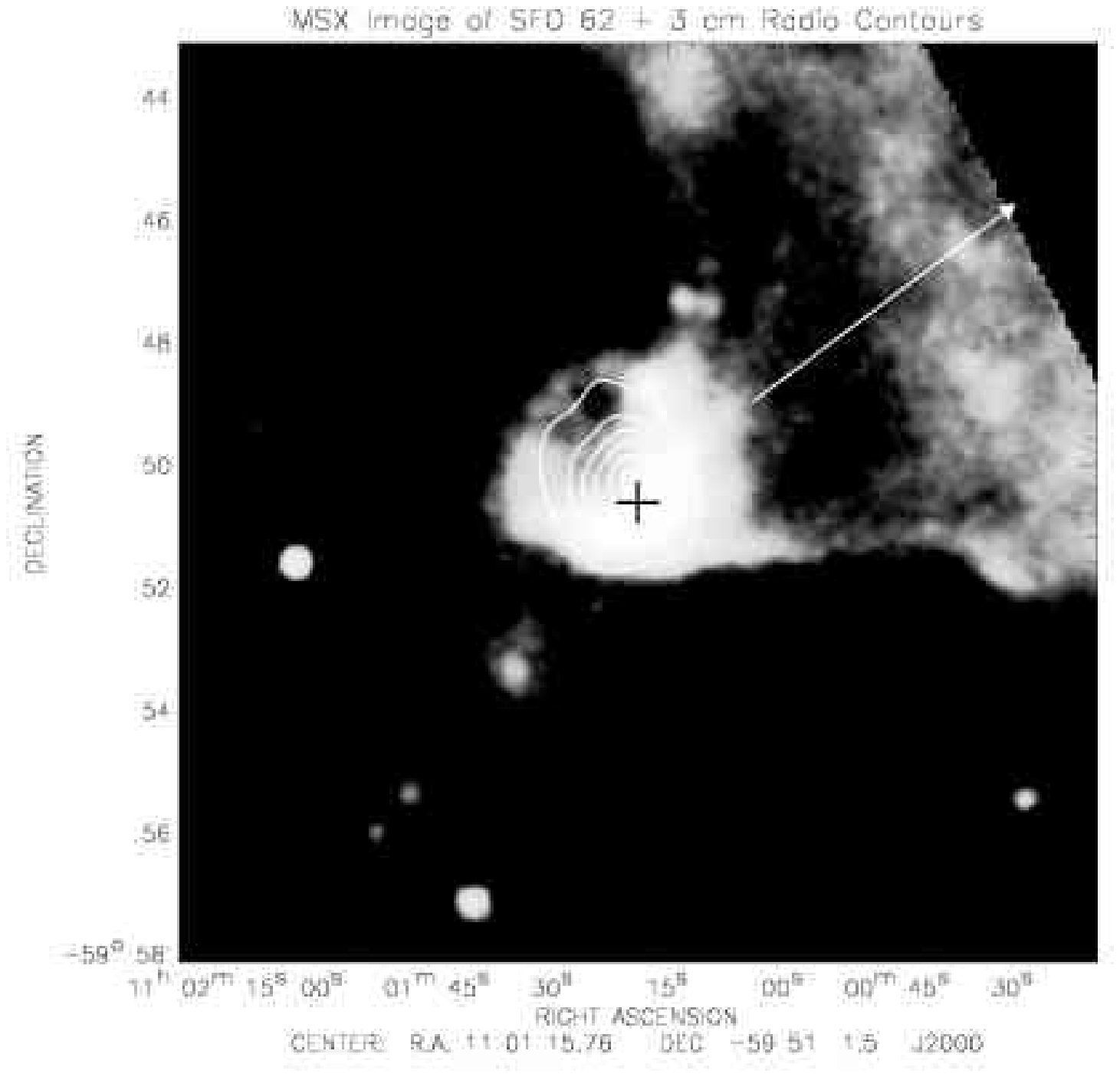}\\
\caption{Example images from the survey of representative Type 1 (SFO 68) and Type 2 (SFO 62) clouds. 
Each cloud is represented
by a pair of images, from the Digitised Sky Survey (\emph{left}) and of the MSX 8 $\mu$m emission
(\emph{right}). Contours of the radio emission are overlaid on each image, along with an
arrow representing the direction towards the suspected ionising star from Yamaguchi et
al.~(\cite{ysmmmof99}) and identifiers in the
case of multiple radio sources. In order to save space, we present only the radio maps 
at the wavelength with
the best resolution or sensitivity for each cloud.}
\label{fig:images}
\end{figure*}

\setcounter{figure}{0}
\begin{figure*}[ht!]
\includegraphics*[scale=0.63,trim=0 0 30 0]{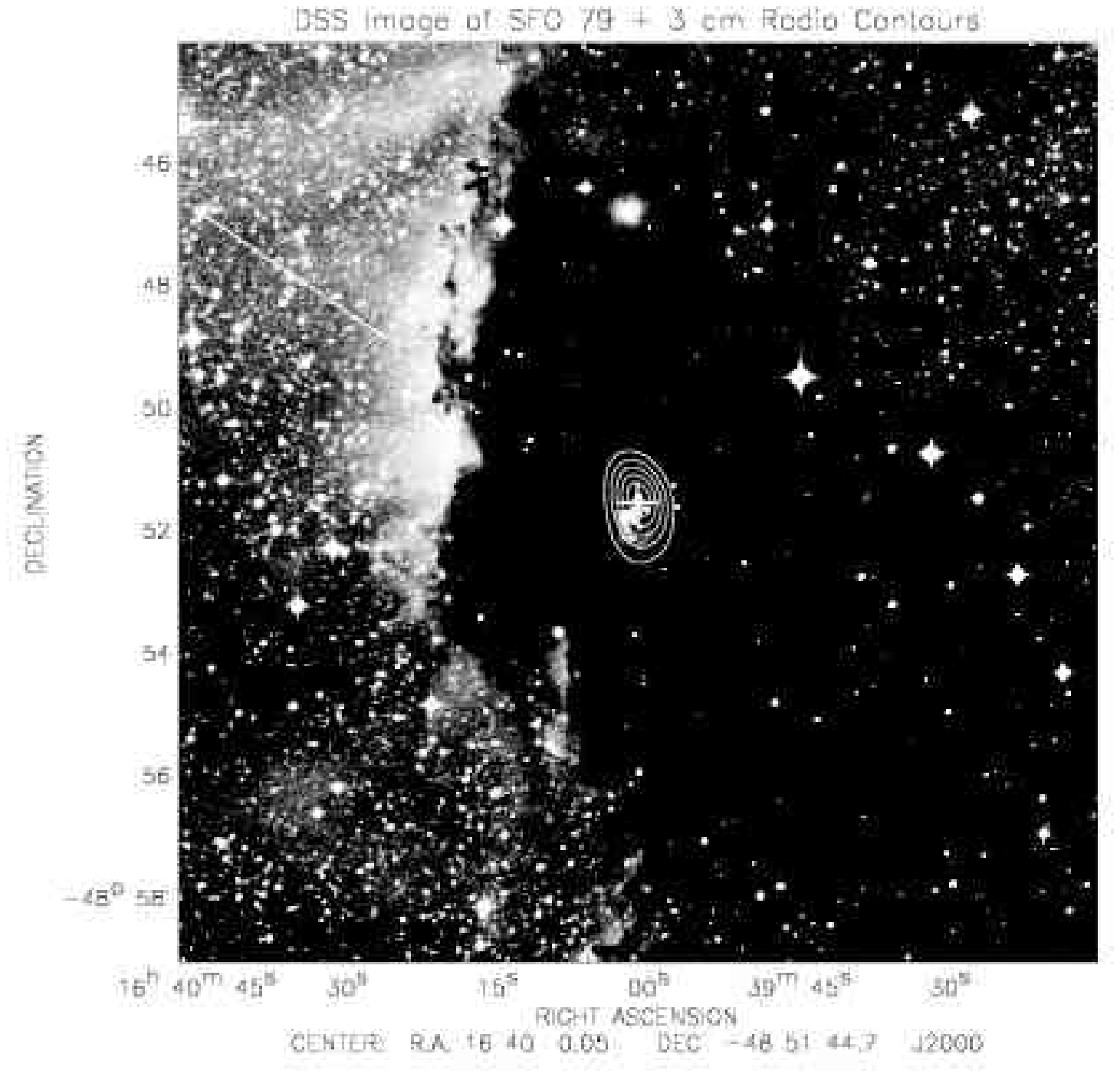}
\includegraphics*[scale=0.63,trim=0 0 30 0]{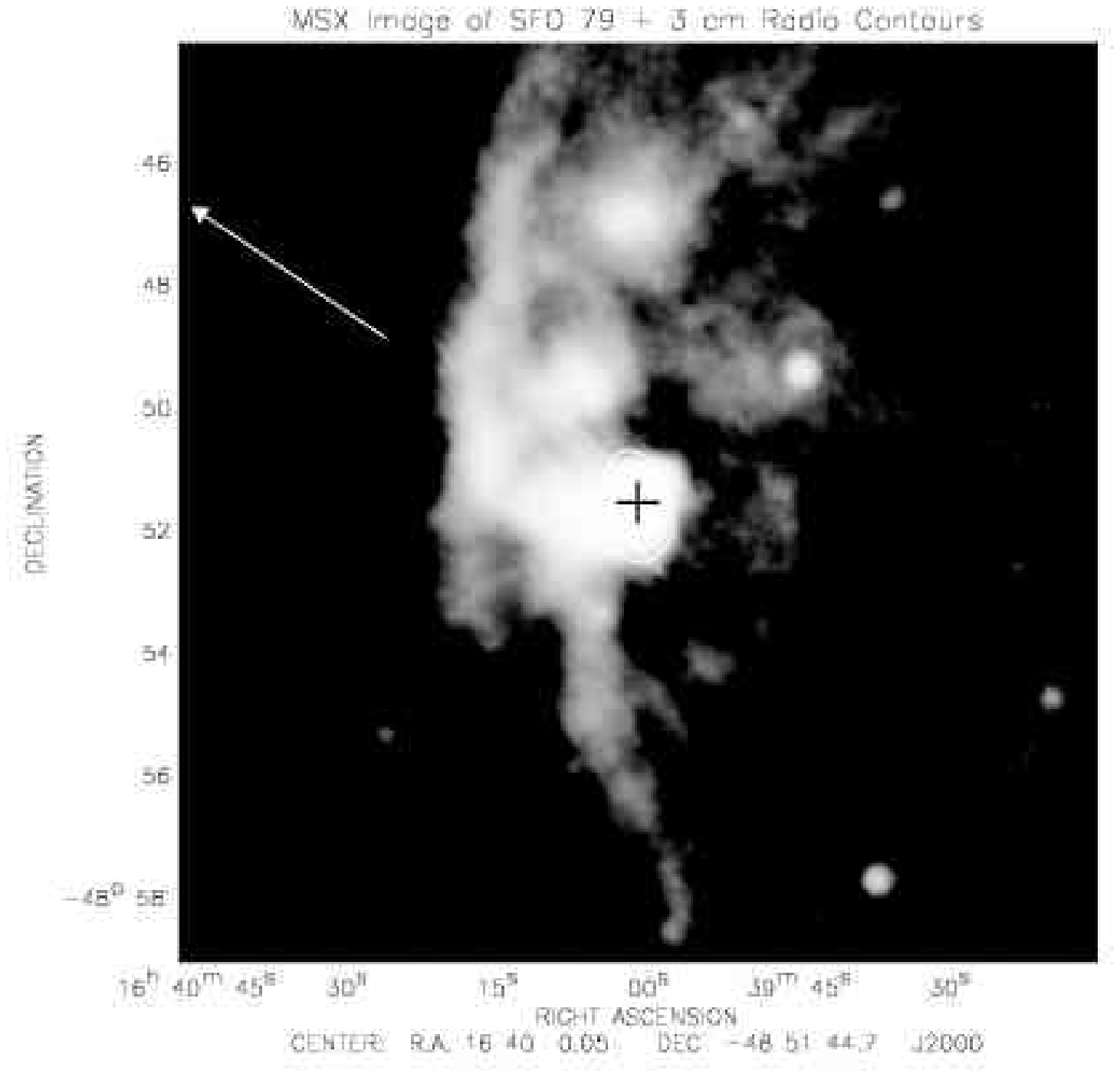}\\
\includegraphics*[scale=0.63,trim=0 0 30 0]{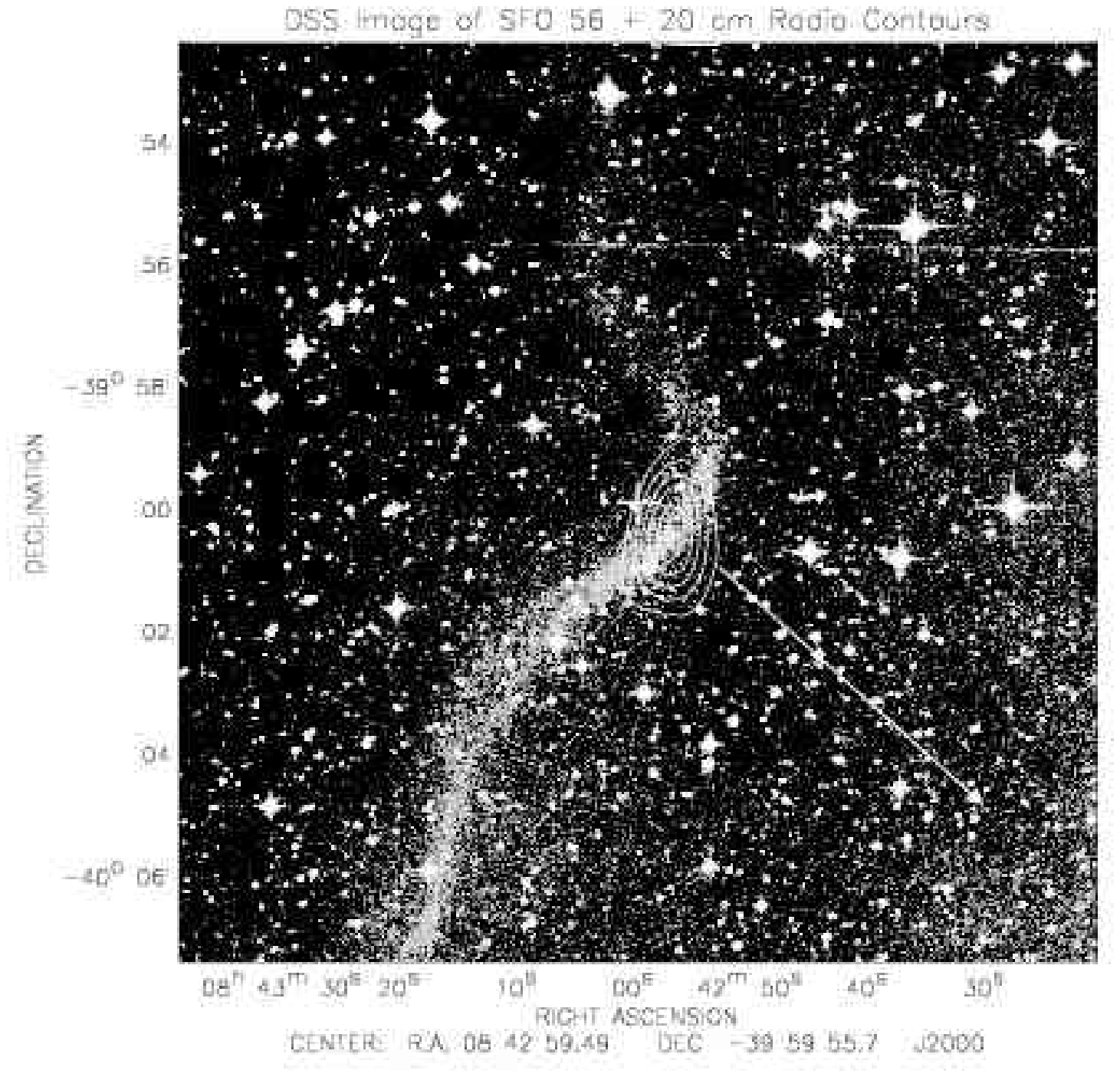}
\includegraphics*[scale=0.63,trim=0 0 30 0]{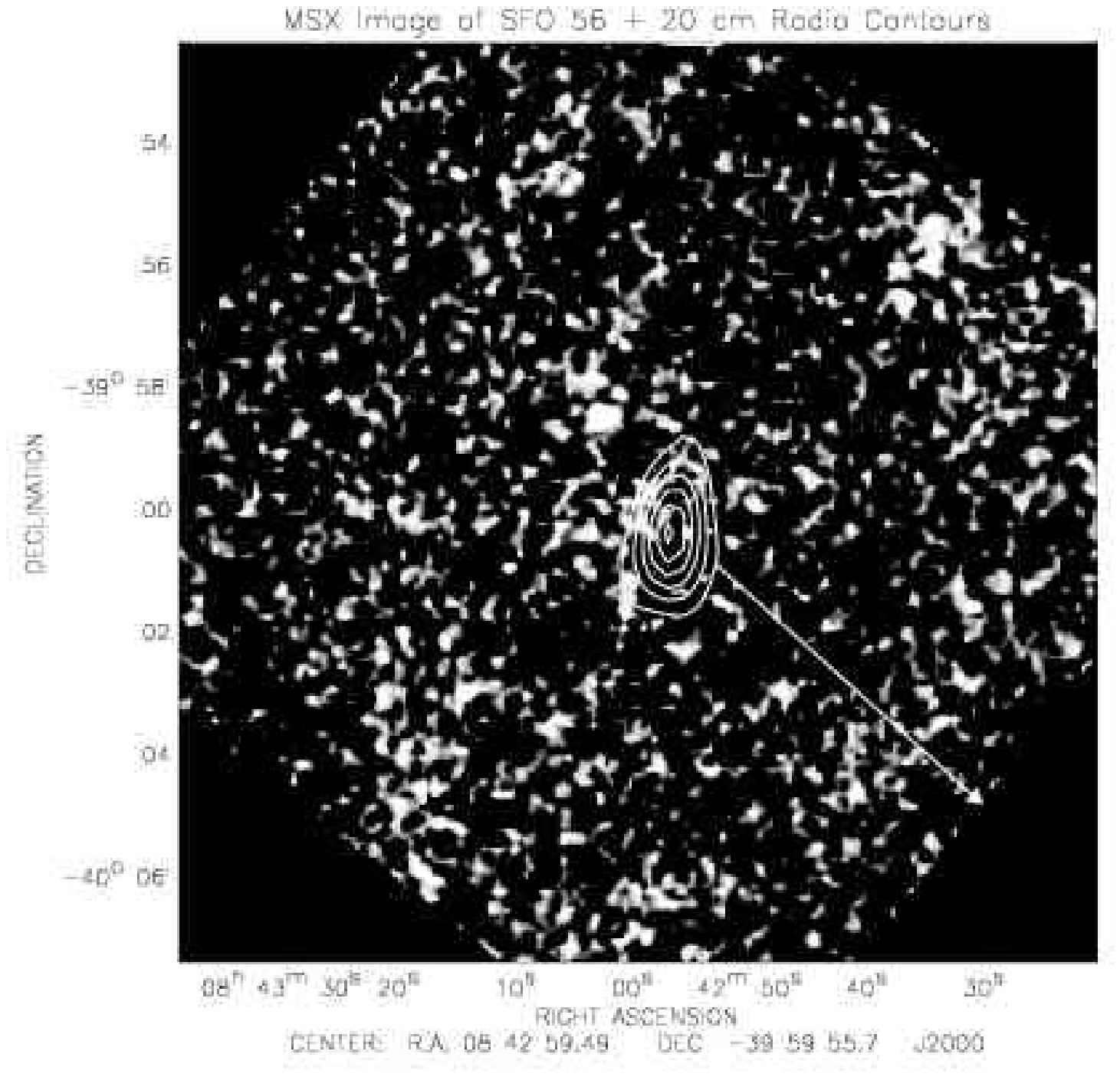}\\
\caption{ continued. Example images from the survey of representative Type 3 (SFO 79) and Type 4 (SFO 56) clouds. 
Each cloud is represented
by a pair of images, from the Digitised Sky Survey (\emph{left}) and of the MSX 8 $\mu$m emission
(\emph{right}). Contours of the radio emission are overlaid on each image, along with an
arrow representing the direction towards the suspected ionising star from Yamaguchi et
al.~(\cite{ysmmmof99}) and identifiers in the
case of multiple radio sources. In order to save space, we present only the radio maps 
at the wavelength with
the best resolution or sensitivity for each cloud.}
\end{figure*}

The radio images  reveal a variety of structures, some displaying  a simple geometry with the
ionised gas traced by the radio emission following the morphology of the optical rim, whereas in
others the emission is concentrated around a central object offset from the optical rim. We have
classified the detected radio sources according to their morphology and association with optical
and MSX 8 $\mu$m emission using the following scheme:

\begin{enumerate}

\item Bright-rim emission clouds with radio and 8 $\mu$m emission positionally coincident with
their bright optical rims. The radio and 8 $\mu$m emission also displays some degree of 
morphological correlation with the cloud rim. These clouds are strong candidates for clouds that
are currently being photoionised by nearby OB star(s) as they display evidence for both
photoionised boundary layers and photodissociation regions that follow the bright optical rim
emission. A total of 18 Type 1 sources were identified in the survey. Examples of the Type 1
rim-emission sources are SFO 58a, SFO 64, SFO 68 and SFO 76. 

\item Broken-rimmed clouds, in which the radio and MSX 8$\mu$m emission is
positionally coincident with the rim of the cloud (as Type 1) but the rim has a
reverse curvature with respect to the normal orientation, i.e.~the rim is curved
towards the molecular cloud, rather than the ionising star. The morphology of
this cloud type resembles the well-known broken cometary globule CG4 in the Gum
Nebula (Reipurth \cite{reipurth83}; Gonz\'alez-Alfonso et al.~\cite{g-acr95}), 
although with the exception that there is a clearly defined
stellar cluster responsible for disrupting the molecular gas. Only a single
cloud of this type (SFO 62) was  identified in the survey.

\item Embedded objects with compact and coincident radio and mid-infrared emission that is
set back from the rim in the heart of the cloud. These objects are strong candidates for
embedded compact HII regions by virtue of the strength of their radio emission and their
infrared colours, which are consistent with the Wood \& Churchwell (\cite{wc89}) IRAS
criteria. There are 4 Type 3 sources identified in the survey, of which SFO 79 is
a good example.

\item Radio emission that is uncorrelated with either the bright optical rim or MSX
8$\mu$m emission. These sources display no obvious morphological or positional
correlation with either the cloud rim or the MSX mid-infrared and are most likely
background extragalactic objects. Twelve radio sources were found to be Type 4 objects,
three times larger than the number of confusing sources predicted from the NVSS source
counts. It is possible that a number of these detections may be due to free-free
emission, but without observations of  higher sensitivity and resolution it is
impossible to classify these sources as such. An example of this is the source
SFO 56, where there is an unresolved radio source positionally coincident with
the cloud rim. We classify this source as Type 4 because there is no supporting
evidence from the MSX 8 $\mu$m emission that the cloud possesses a PDR and that
the radio emission is associated with an ionised boundary layer. 

\end{enumerate}

MSX 8 $\mu$m and Digitised Sky Survey red images of representative sources overlaid with radio
contours illustrating the classification scheme can be found in Fig.~\ref{fig:images}. Images
of the remaining clouds are available only in the online supplement to this paper.

\subsection{Ionised boundary layers associated with BRCs}
\label{sect:ibl_derivn}

For those clouds with associated radio emission that is identified as Type 1, i.e.~where the
emission is highly likely to be free-free emission from the rim of the cloud, we 
evaluate the ionising photon flux impinging upon the cloud and the electron 
density and pressure of the ionised boundary layer. In order to determine these quantities
we use the general equations from Lefloch et al.~(\cite{llc97}). Rearranging
their Equation 1, the ionising photon flux $\Phi$ arriving at the cloud rim may be written
in units of cm$^{-2}$\,s$^{_-1}$ as

\begin{equation}
\label{eqn:phi}
\Phi = 1.24 \times 10^{10} \,S_{\nu}\, T_{\rm e}^{\,0.35}\, \nu^{0.1} \,\theta^{-2}, 
\end{equation}

where $S_{\nu}$ is the integrated radio flux in mJy, $T_{\rm e}$ is the effective
electron temperature of the ionised gas in K, $\nu$ is the frequency of the free-free
emission in GHz and $\theta$ is the angular diameter over which the emission is
integrated in arcseconds.

The electron density ($n_{\rm e}$) of the ionised boundary layer surrounding the cloud may also
be derived from the integrated radio flux $S_{\nu}$ by subsituting for the ionising photon flux
in Equation 6 of Lefloch et al.~(\cite{llc97}). The electron density in cm$^{-3}$
is given by:

\begin{equation}
\label{eqn:ne}
n_{\rm e} = 122.41 \,\,\sqrt{\frac{S_{\nu}\, T_{\rm e}^{0.35}\, \nu^{0.1}\, \theta^{-2}}{\eta R}},
\end{equation}

where those quantities common to both Eq.~(\ref{eqn:phi}) and (\ref{eqn:ne}) are in the same
units, $R$ is the radius of the cloud in pc and  $\eta$ is the effective thickness of the
ionised boundary layer as a fraction of the cloud radius. Bertoldi (\cite{bertoldi}) shows that
the effective thickness of the IBL $\eta$ varies over the range 0.1--0.2 of the cloud radius and
is mainly dependent upon the product of the ionising flux and the cloud curvature. Here, we
assume $\eta=0.2$, which implies that the derived electron density is strictly a lower limit
(albeit by no more than a factor of $\sqrt{2}$). As several of the clouds are unresolved by our radio
observations we estimated the cloud radii ($R$) from the Digitised Sky
Survey images of the cloud rims published in Sugitani \& Ogura (\cite{so94}). The pressure
within the ionised boundary layer may be evaluated from the electron density $n_{\rm e}$ and
temperature $T_{\rm e}$ via:

\begin{equation}
\label{eqn:pi}
 P_{i} = 2n_{\rm e}\,T_{\rm e}\,k. 
\end{equation}

 As the pressure within the IBL depends upon the electron density, the pressure is again a strict
lower limit to within a factor of $\sqrt{2}$. Values for $\Phi$, $n_{e}$ and $P_{i}$ were calculated
using Eq.~(\ref{eqn:phi})--(\ref{eqn:pi}), assuming a boundary layer thickness of $\eta = 0.2$
and an  effective electron temperature of $T_{\rm e} = 10^{4}$ K.  In order to derive a global
average of $\Phi$, $n_{e}$ and $P_{i}$ for each cloud the source-integrated fluxes contained in
Tables \ref{tbl:6/3sources} and \ref{tbl:20/13sources} were used. Where the sources were
detected at more than one wavelength (i.e. 20+13 cm or 6+3 cm), separate values were calculated
at each wavelength and found to agree closely (to within the measurement errors). The separate
wavelength values for $\Phi$, $n_{e}$ and $P_{i}$ were then averaged together for each cloud and
are given in Table \ref{tbl:ibl_results}.

\begin{table*}
\begin{center}
\caption{Values for the measured ionising flux, predicted
ionising flux, the
measured electron density and ionised gas pressure for radio sources detected in the
survey.}
\label{tbl:ibl_results}
\begin{tabular}{lccccc}\hline\hline
 & Cloud Radius & Measured ionising flux & Predicted ionising flux & Electron Density & Ionised gas
 pressure \\
Source ID & $R$ (pc) &  $\Phi$ (10$^{8}$ cm$^{-2}$\,s$^{-1}$) &$\Phi_{P}$ (10$^{8}$
 cm$^{-2}$\,s$^{-1}$)
 & $n_{\rm e}$ (cm$^{-3}$ )& $P_{\rm i}/k_{B}$ (cm$^{-3}$\,K)\\ \hline
SFO 57 & 0.10 & 7.7 & 29 & 214 & 4.3\,\,10$^{6}$ \\
SFO 58b & 0.18 & 1.5 & 15 & 72 & 2.8\,\,10$^{5}$ \\
SFO 61a & 0.81 & 3.1 & 0.2 & 85 & 1.9\,\,10$^{5}$ \\
SFO 63 & 0.57 & 2.1 & 0.2 & 147 & 2.9\,\,10$^{6}$ \\
SFO 64 & 0.75 & 2.3 & 20 & 133 & 2.7\,\,10$^{6}$ \\
SFO 66a & 0.16 & 2.4 & 6.4 & 93 & 1.9\,\,10$^{6}$ \\
SFO 66b & 0.16 & 2.7 & 6.4 & 98 & 2.0\,\,10$^{6}$ \\
SFO 67a & 0.35 & 3.0 & 22 & 72 & 1.4\,\,10$^{6}$ \\
SFO 68 & 0.38 & 15 & 63 & 153 & 3.1\,\,10$^{6}$ \\
SFO 75 & 0.22 & 200 & 570 & 730 & 1.5\,\,10$^{7}$ \\
SFO 76 & 0.12 & 22 & 110 & 333 & 6.7\,\,10$^{6}$ \\
SFO 77 & 0.002 & 9.4 & 1.6 & 1660 & 3.3\,\,10$^{7}$ \\
SFO 78 &  0.002 & 1.3 & 0.5 & 533 & 1.1\,\,10$^{7}$ \\
SFO 82b & 0.24 & 4.9 & 1.0 & 111 & 2.2\,\,10$^{6}$ \\
SFO 83 & 0.19 & 3.7 & 1.9 & 108 & 2.2\,\,10$^{6}$ \\
SFO 84 & 0.40 & 5.4 & 2.3 & 89 & 1.8\,\,10$^{6}$ \\
SFO 89 & 0.10 & 16 & 19 & 303 & 6.1\,\,10$^{6}$ \\ \hline
\end{tabular}
\end{center}
\end{table*}

It is also instructive to compare the value for the ionising flux $\Phi$ that is
derived from the measured free-free radio flux density to the value that can be
\emph{predicted} from the spectral type of the suspected ionising star
($\Phi_{P}$). Contrasting the observed ionising flux $\Phi$ to the predicted
ionising flux $\Phi_{P}$ allows us to investigate the likelihood that  the
clouds are being ionised by the suspected star in question and determine whether
any additional ionising sources may be present or if the suspected ionising star
is located much further from the cloud than its projected distance on the sky
suggests.

We identified the suspected ionising stars for the BRCs in our sample by using a combination of
those identified by Yamaguchi et al.~(\cite{ysmmmof99}) and a literature search carried out
using the SIMBAD database of astronomical catalogues (\texttt{http://simbad.u-strasbg.fr}). Two
BRCs are missing from the list in Yamaguchi et al.~(\cite{ysmmmof99}), SFO 77 and SFO 78. For
these two stars we assume that they are primarily ionised by the nearby B1 III star $\sigma$ Sco
(Sugitani \& Ogura \cite{so94}), also known as HD 147165, and the B1 star CCDM J16212$-$2536AB.
We also note that there is a discrepancy regarding the  distance to $\sigma$ Sco; Sugitani \&
Ogura (\cite{so94}) quote a distance of 1.65 kpc to $\sigma$ Sco, whereas this star appears to
be a member of the Sco OB2 complex which is at a distance of $\sim$ 150 pc (Eggen
\cite{eggen98}). In this paper we assume a distance of 150 pc to $\sigma$ Sco and its two
associated BRCs.

\begin{table*}
\begin{center}
\caption{The HII regions and ionising stars of the bright-rimmed clouds in our
study. Parentheses indicate that the listed object is a star rather than an HII
region. The ionising star, spectral type and assumed distance values are drawn from
Yamaguchi et al.~(\cite{ysmmmof99}) apart from $\sigma$ Sco where the spectral
type and assumed distance are taken from Eggen (\cite{eggen98}) and 
HD 305938, CCDM J16212$-$2536AB, HD 326286 and  HD 152245, whose values are drawn from SIMBAD.}
\label{tbl:ionising_stars}
\begin{tabular}{lllll}\hline\hline
HII region & Ionising star(s) & Spectral Type & $D$ (kpc) & Associated BRCs
\\\hline
S 306 & MFJ 3 & B0 V & 3.6 & SFO 47\\
S 307 & LSS 458 & O5 & 4.2 & SFO 49\\
 & LSS 467 & O9.5 III & & \\
RCW 14 & HD 57236 & O8 V & 3.9 & SFO 45 \\
Gum Nebula & $\zeta$ Pup & O4 If & 0.45 & SFO 46, SFO 48, SFO 50--53\\
 & $\gamma^{2}$ Vel & WC 8+O8 III & & \\
NGC 2626 & vBH 17a & B1 V & 0.95 & SFO 54\\
RCW 27 & HD 73882 & O8 V & 1.2& SFO 55, SFO 56\\
RCW 32 & HD 74804 & B0 V & 0.70 & SFO 57, SFO 58\\
RCW 38 & RCW 38 IRS 2 & O5 & 1.7 & SFO 59, SFO 60\\
NGC 3503 & vBH 46a & B0 Ve & 2.9 & SFO 61--63\\
 & HD 305938 & B1 V & & \\
BBW 347 & LSS 2231 & B0 V & 2.7 & SFO 64\\
RCW 62 & HD 101131 & O6.5 N & 1.4 & SFO 65--70\\
 & HD 101205 & O6.5 & & \\
 & HD 101436 & O7.5 & & \\
(Cen R 1) & vBH 59 & B6 I & 2.0 & SFO 71\\
RCW 75 & HD 115455 & O7.5 III & 1.9 & SFO 72, SFO 73\\
RCW 85 & HD 124314 & O6 V & 1.5 & SFO 74\\
RCW 98 & LSS 3423 & O9.5 IV & 2.8 & SFO 75\\
RCW 105 & HD 144918 & O7 & 1.8 & SFO 76\\
($\sigma$ Sco) & HD 147165 & B1 III & 0.15 & SFO 77, SFO 78 \\
 & CCDM J16212$-$2536AB & B1 & & \\
RCW 108 & HD 150136 & O5 III & 1.4 & SFO 79--81\\
 & HD 150135 & O6.5 V & & \\
RCW 113/116 & HD 152233 & O6 III & 1.9 & SFO 82--85\\
 & HD 326286 & B0 & & \\
 & HD 152245 & B0 Ib & & \\
RCW 134 & HD 161853 & O8 V & 1.6 & SFO 86\\
M 8 & HD 164794 & O4 V & 1.9 & SFO 87, SFO 88\\
S 29 & HD 165921 & O7.5 V & 1.9 & SFO 89\\ \hline
\end{tabular}
\end{center}
\end{table*}

We estimate the predicted ionising photon flux ($\Phi_{P}$) illuminating the Type 1 BRCs from
the tables of Panagia (\cite{panagia}) and the projected distance between the star(s) and each
cloud. $\Phi_{P}$ is a strict upper limit to the ionising flux illuminating the cloud, due to
the fact that the  projected star-cloud distance represents a lower limit to the actual distance
between the ionising star and the cloud.  The values of $\Phi_{P}$ that we determined by this
approach are listed in Table \ref{tbl:ibl_results} alongside the values of $\Phi$ derived from
Eq.~(\ref{eqn:phi}) and the measured radio flux densities. In most cases it can be seen that
$\Phi_{P} > \Phi$ to within at least a factor of 10 as is expected.

Several clouds do not, however, follow this trend (notably SFO 61a, SFO 63, SFO 77, SFO 78, SFO
82b, SFO 83 and SFO 84). The most likely reason behind this is that the ionising stars we have
identified are not those responsible for exciting the rims of the BRCs or that their spectral
classifications are incorrect (even a misclassification by only half a spectral class can
increase or decrease the predicted Lyman flux by a factor of 2).  Further investigation and
classification of the stars exciting the HII regions is required, perhaps including optical
spectroscopy in order to determine the stellar spectral types independent of any reddening
effects.

\subsection{Upper limits to the ionising flux for non-detected clouds}
\label{sect:uplim}

For those clouds that do not exhibit any associated radio emission we were able
to derive an upper limit ($\Phi_{\rm max}$) to the ionising flux illuminating
their surfaces by substituting a 3$\sigma$ upper limit for the radio flux
density into Eq.~(\ref{eqn:phi}). The wavelength used in the derivation of
the ionising flux upper limit was typically the longest of the two wavelengths at which each
cloud was observed, as the larger beams of the longer wavelengths allow tighter
constraints to be placed upon $\Phi_{\rm max}$. Values of the 3$\sigma$ flux
limit and the wavelength used in the derivation of  the  observed upper limit to
the ionising flux $\Phi_{\rm max}$ are given in Table \ref{tbl:uplim}. 

\begin{table*}
\begin{center}
\caption{Upper limits to the observed ionising flux illuminating clouds with no
associated radio emission. The predicted ionising fluxes are  calculated
from the spectral type and projected distance of the ionising stars contained in Table
\ref{tbl:ionising_stars}.}
\label{tbl:uplim}
\begin{tabular}{lccccc}
\hline
\hline
Source ID & $\lambda$ & 3$\sigma$ flux upper limit & Cloud Radius & Max.~observed ionising flux &
Predicted ionising flux \\
 & (cm) & $S_{\rm max}$ (mJy) & $R$ (pc) & $\Phi_{\rm max}$ (10$^{6}$ cm$^{-2}$\,s$^{-1}$) &
$\Phi_{P}$ (10$^{6}$ cm$^{-2}$\,s$^{-1}$) \\
\hline
SFO 45 & 20 &    5.16 &  0.30 &  24 &	  38 \\
SFO 46 & 13 &    0.75 &  0.05 &  18 &	  13 \\
SFO 48 & 20 &    2.58 &  0.01 &  20 &	  15 \\
SFO 50 & 20 &    3.60 &  0.05 &  21 &	  41 \\
SFO 51 & 20 &    2.82 &  0.15 &  19 &	  69 \\
SFO 52 & 20 &    9.45 &  0.20 &  78 &	  8.6\\
SFO 53 & 20 &    9.60 &  0.10 &  84 &	  9.0 \\
SFO 54 & 20 &    5.10 &  0.21 &  36 &	  0.6 \\
SFO 55 & 20 &    1.50 &  0.48 &  11 &	  23 \\
SFO 60 & 20 &  360.00 &  0.12 & 2900 &   2500 \\
SFO 71 & 13 &    4.10 &  0.61 &  140 &     0.06 \\
SFO 72 & 20 &   10.80 &  0.54 &  92 &	  820 \\
SFO 73 & 13 &    3.00 &  0.27 &  100 &     190 \\
SFO 80 & 20 &   30.00 &  0.28 &  220 &    230 \\
SFO 81 & 13 &    6.90 &  0.19 &  190 &     270 \\
SFO 86 & 13 &    3.00 &  0.36 &  56 &	  1600 \\
SFO 87 & 20 &  126.00 &  0.43 &  500 &     1300 \\
SFO 88 & 20 &   78.00 &  0.40 &  300 &      990 \\
\hline
\end{tabular}
\end{center}
\end{table*}

We again compare the upper limit to the ionising flux  derived from the
observations to that predicted from the spectral type and location of the
suspected ionising star given by Yamaguchi et al.~(\cite{ysmmmof99}). The
predicted ionising flux illuminating each cloud  ($\Phi_{P}$) was evaluated
following the procedure outlined in Sect.~\ref{sect:ibl_derivn} and using the
stars and spectral types contained in Table \ref{tbl:ionising_stars}. 

Comparing the values of $\Phi_{\rm max}$ to $\Phi_{P}$ contained in Table
\ref{tbl:uplim} reveals that for roughly half of the clouds (8 clouds in total)
the lack of detectable radio emission is fully consistent with the predicted
ionising flux. For these clouds our observations show that the lack of
detectable radio free-free emission from the ionised boundary layer is
consistent with the ionising fluxes emitted from the suspected ionising stars
identified in Table \ref{tbl:ionising_stars}. The sensitivity of our
observations was ultimately restricted by the relatively limited $uv$-plane
coverage that we could devote to each cloud and confusing flux introduced by
sidelobes from nearby bright objects. Deeper observations with better $uv$-plane
coverage are required to detect any free-free emission from the ionised boundary
layers in these clouds.

The remaining eleven clouds listed in Table \ref{tbl:uplim} have predicted
ionising fluxes that are inconsistent with the upper limits derived from our
observations (i.e.~$\Phi_{P} > \Phi_{\rm max}$).  For the majority of the clouds
this is more than likely due to the differences between the true star-cloud
distance and the projected distance that we measure on the sky, and/or the
negligible UV absorption assumptions inherent in our calculation of the
predicted ionising flux $\Phi_{P}$. For example, if the star-cloud vector is
inclined by 60\degr\ to the line-of-sight, then the true distance between the
ionising star and the bright-rimmed cloud is increased by a a factor of 15\% and
the predicted ionising flux will decrease by 30\%. 

Changes in the predicted ionising flux of a factor of a few may be plausibly
explained by the projected distance between the ionising star and the
bright-rimmed cloud, but there are two clouds in Table \ref{tbl:uplim} (SFO 72
and SFO 86) with predicted ionising fluxes that are roughly an order of
magnitude greater than the maximum ionising flux determined from the survey
observations. It is unlikely that the disparity in predicted and observed
ionising flux limits can be explained by geometric effects alone. We will investigate these two
clouds in more detail in Sect.~\ref{sect:sfo72_86}.

\subsection{Embedded radio sources -- compact HII regions?}
\label{sect:embedded}

\begin{table*}
\begin{center}
\caption{Infrared and radio-derived spectral types for the Type 3 and Type 2
radio sources detected in the survey. Parentheses indicate an upper limit to the
infrared luminosity}
\label{tbl:type3}
\begin{tabular}{ccccccc}
\hline
\hline
Source Id        & IRAS PSC ID  &  IR Luminosity & Spectral type  & Flux density
& Ionising photon flux &  Spectral type  \\
        &   & ($L_{IR}$/$L_{\odot}$)	 & (IR)		& $S_{\nu}$ (mJy) 	& Log (\emph{N$_{i}$}) & (Radio)\\
\hline
SFO 59 & 08563--4711 & 13000  &  B0.5 & 10400 & 48.4 & O8\\
SFO 62 & 10591--5934 & 18000 &    B0--B0.5 &126 & 47.3 & B0\\
SFO 74 & 14159--6111 & 5500  &    B1 & 26.3  & 45.5 & B1\\
SFO 79 & 16362--2501 & (4400)  &  B1 & 4800  & 47.9 & O9.5\\
SFO 85a & 16555--4237 & 18000  &    B0.5 &  40.3& 45.9& B0.5--B1 \\
\hline\hline
\end{tabular}
\end{center}
\end{table*}

Four Type 3 radio sources were detected in the survey: SFO 59, SFO 74, SFO 79
and SFO 85a. These sources are strong (with radio fluxes of several tens of mJy
to over a Jy), mostly unresolved, displaced from the bright cloud rim and associated
with compact MSX mid-infrared emission. The positions of the unresolved radio
emission are coincident with the IRAS and MSX point source positions.  The
characteristics of these sources are similar to compact or ultracompact HII
regions, i.e.~they are infrared luminous ($L_{IR} \sim$ 10$^{3}$--10$^{4}$
\lsol) compact radio sources associated with molecular clouds (and possibly
deeply embedded within the clouds). In addition the IRAS colours of these
sources meet the criteria proposed by Wood \& Churchwell (\cite{wc89}) for
ultracompact HII regions, namely
that $\log F_{60}/\log F_{12} \ge 1.3$ and $\log F_{25}/\log F_{12} \ge 0.57$. It
is possible that these sources may be compact or ultracompact HII regions
embedded within the BRCs.

In order to investigate the physical nature of these sources we have calculated
their far-infrared (FIR) and radio luminosities so that we may determine the
spectral class of any potential embedded massive Young Stellar Objects (YSOs).
The FIR luminosity from an embedded YSO is almost entirely  due to the
processing of absorbed stellar radiation from circumstellar dust and so the FIR
luminosity can be used to estimate the bolometric luminosity and  spectral class
of the embedded YSO. 

We assume that the FIR luminosity arises from a single embedded star and
estimate the luminosity from the integrated IRAS fluxes. Given the rather coarse
scale  of the IRAS FWHM beam  this is an unrealistic assumption and the emission
may arise from an embedded cluster instead of a single object. Nevertheless,
Wood \& Churchwell (\cite{wc89}) showed that for a realistic initial mass function
estimate the spectral type of the most massive member in a cluster is only
1.5--2 spectral classes lower than that derived for the single embedded star
case. The resulting FIR luminosity was corrected following the method of
Sugitani \& Ogura (\cite{so94}) to take into  account contributions from the
flux longwards of 100 $\mu$m. Distance estimates to the embedded sources were
also taken from Sugitani \& Ogura (\cite{so94}).

It is also possible to independently derive the spectral type from the observed
radio flux. In this case the radio flux is due to free-free emission originating
from an embedded compact or ultracompact HII region. If the HII region is
assumed to be optically thin at radio wavelengths and in photoionisation
equilibrium (i.e.~the number of photons ionising new material is balanced by the
recombination rate) then the integrated radio flux may be related to the total
number of ionising photons emitted from the star powering the HII region (e.g.~Wood \& Churchwell
\cite{wc89}; Carpenter et al.~\cite{css90}). The tables of Panagia
(\cite{panagia}) can then used to convert the ionising photon flux into a
spectral type.

In order to derive the ionising photon flux $N_{i}$ emitted from the massive
YSOs we used Eq.~(7) from Carpenter et al.~(\cite{css90})
relating $N_{i}$ to the radio flux density, i.e.

\begin{equation}
N_{i}=7.7\times10^{43}\,\,S_{\nu}\,\, D^{2} \,\, \nu^{0.1}
\label{eqn:ni}
\end{equation}

where $N_{i}$ is measured in units of photons\,\,s$^{-1}$, $S_{\nu}$ is the
integrated radio flux density of the embedded radio source in mJy, $D$ is the
distance to the BRC in kpc and $\nu$ is the frequency at which the integrated
flux density is determined. Note that we have removed the 5 GHz term from
Eq.~(7) of Carpenter et al.~(\cite{css90}) and adjusted the equation
coefficient accordingly.

The radio and infrared derived spectral types of the four Type 3 sources in
our survey are listed in Table \ref{tbl:type3}. To this list we also add SFO 62,
which is the only Type 2 broken-rimmed cloud in the survey. We included SFO 62 as
it seems likely that it is a more evolved YSO cluster on the verge of breaking
free from its natal molecular cloud (see Sect.~\ref{sect:sourceid} \&
Fig.~\ref{fig:images}). It is useful to compare the physical properties of the
cluster associated with SFO 62 to those of its presumably less evolved cousins.

As can be seen from Table \ref{tbl:type3} all of the Type 3 sources are
associated with either late O-type stars or early B-type stars. It is thus
highly likely that these sources are massive YSOs surrounded by compact or
ultracompact HII regions and embedded within the molecular material of the
bright-rimmed clouds. The infrared and radio-derived spectral types agree
surprisingly well, which may indicate that the UV radiation within the embedded
compact HII region is not significantly absorbed by dust within the ionised gas
(Wood \& Churchwell \cite{wc89}). In order to confirm this hypothesis
sub-millimetre or millimetre-wave  continuum observations are required to
constrain the spectral energy distribution longward of 100 $\mu$m and confirm
our estimate of the IRAS luminosity correction factor. We will discuss the
nature of the embedded sources further in Sect.~\ref{sect:discuss}.

\section{Discussion}
\label{sect:discuss}

\subsection{Are the BRCs in pressure equilibrium?}

A total of 18 clouds were identified as possessing detectable radio emission
from their ionised boundary layers and classified as Type 1 radio sources. In
this section we will examine the physical properties of the ionised boundary
layers associated with these clouds and estimate the likely pressure balance
between the ionised boundary layer (IBL) and the interior molecular gas of the
BRC. The pressure balance between the ionised and neutral components of BRCs has
been revealed as a sensitive diagnostic of their status (Lefloch \& Lazareff
\cite{ll94}). The photoionisation-induced shocks that are implicated as
potential star-formation triggers (e.g.~Elmegreen \cite{e91}; Sugitani et al.~\cite{sfo}; 
Sugitani et al.~\cite{sfmo89}) are restricted from propagating
into the clouds if the pressure in the ionised boundary layer is less than that
in the interior neutral gas of the cloud. The pressure imbalance causes the
photo-ionisation shocks to stall at the surface of the cloud (Lefloch \&
Lazareff \cite{ll94}).

Radiative-driven implosion (RDI) modelling of BRCs shows that the evolution of
the cloud is mainly dependent  upon the duration of its ionisation (Lefloch \&
Lazareff \cite{ll94}; Bertoldi \cite{bertoldi}). Due to the increasing 
recombination in the ionised boundary layer the pressure of the layer rises
over time and clouds that were initially over-pressured will eventually  reach
equilibrium. This feature of BRC evolution allows us to separate out those
clouds that are potential regions of induced star formation for further study.
If the clouds are currently overpressured with respect to their IBL then it is
unlikely that photoionisation shocks are propagating (or have propagated)
through the molecular gas of the clouds. It is thus doubtful that any existing
star formation within these  over-pressured clouds could have been caused by
the photoionisation of the clouds. Conversely, if the clouds are underpressured
with respect to their IBL then it is highly likely that photoionisation-induced
shocks are propagating into the cloud and this raises the possibility that the
star formation within the clouds could have been induced by the action of the
nearby OB star or stars.

We have determined the pressure in the ionised boundary layers from our radio
continuum observations for the 18 BRCs classed as Type 1 radio sources. Values
for the ionised gas pressure over Boltzmann's constant $P_{i}/k$ range from 2.8
$\times$ 10$^{5}$ to 3.3 $\times$ 10$^{7}$ cm$^{-3}$\,K, with a median value of
a few 10$^{6}$ cm $^{-3}$\,K.  We estimate that, due to a combination of measurement errors and
uncertainities in the relative IBL thickness ($\eta$), the pressure of the ionised gas in
underestimated by no more than a factor of two. In order to evaluate the pressure of the neutral
molecular gas in the interior of the BRCs we require molecular line data in
order to constrain the velocity dispersion of the molecular gas from the width
of molecular rotation lines and the density of the molecular gas from the
integrated intensity of the emission. 

The overall molecular pressure is
comprised of both thermal and turbulent contributions, but under typical
molecular cloud conditions the turbulent pressure component dominates (as
evidenced by suprathermal linewidths; e.g.~Larson \cite{larson81}).
Neglecting the thermal pressure component, the molecular pressure $P_{m}$ may be
written as $P_{m} \simeq \sigma^{2}\,\rho_{m}$, where $\sigma$ is the velocity
dispersion  and $\rho_{m}$ is the density of the gas. The velocity dispersion can
be expressed in terms of the observed FWHM linewidth $\Delta v$ as $\sigma^{2} =
\left< \Delta v\right>^{2}/8 \ln 2$.

There are few existing molecular line observations of these clouds and currently
the most comprehensive study is that of Yamaguchi et al.~(\cite{ysmmmof99}), who
carried out wide-field mapping of the HII regions from Sugitani \& Ogura
(\cite{so94}) using the NANTEN 4m telescope.  We originally attempted to
determine the density and velocity dispersion for each of the 18 clouds
possessing measurable ionised boundary layers by using the $^{13}$CO
observations of Yamaguchi et al.~(\cite{ysmmmof99}). However the combination of
coarse resolution and sensitivity meant that not all of the clouds in Table
\ref{tbl:ibl_results} were detected (of the 18 clouds in our sample Yamaguchi et
al.~(\cite{ysmmmof99}) detected only 8). The maps of these clouds have an
intrinsic resolution of 4\arcmin\, due to non-Nyquist sampling of the NANTEN
2.7\arcmin\ beam, and so almost all of the clouds  in our sample
were unresolved and/or blended with nearby emission.

With the existing data it is not possible to determine the individual
properties of each cloud.  Nevertheless we have used the results from Yamaguchi
et al.~(\cite{ysmmmof99}) to derive a global mean of the density and velocity
dispersion, which we apply to our entire sample of BRCs from Table
\ref{tbl:ibl_results}. Whilst we cannot definitively prove that individual 
clouds are under- or over-pressured using this approach, we can nevertheless 
identify those clouds in our sample with ionised boundary layer pressures that
deviate significantly from the mean molecular pressure. These clouds are then
the likeliest candidates to be in pressure imbalance.

The mean FWHM linewidth of the 8 clouds detected by Yamaguchi et al.~is 3.5
km\,s$^{-1}$, which compares favourably to the linewidths of other BRCs that
have been observed in $^{13}$CO or other molecular lines  and which typically
range between 1 and 3 km\,s$^{-1}$ (e.g.~Thompson et al.~\cite{thompson03a};
Codella et al.~\cite{cbnst01}; Lefloch et al.~\cite{llc97}). The 
mean H$_{2}$ number density of the clouds is 220 cm$^{-3}$, which is about a
factor of 50 less than typical star-forming regions  (Evans et
al.~\cite{evans99}) and other bright-rimmed clouds (Thompson et
al.~\cite{thompson03a}). Because Yamaguchi et al.~only mapped the clouds in the
ground state transition of $^{13}$CO (with some limited $^{12}$CO
single-position spectra) it is not possible to accurately estimate  the optical
depth of $^{13}$CO. In addition the low critical density and dipole moment of
the CO molecule mean that emission from low-density, low-temperature gas is
preferentially sampled. As a result the mean density that we derive from the 
results of Yamaguchi et al.~(\cite{ysmmmof99}) is likely to underestimate the
true density within the clouds.

From the mean linewidth and density we evaluate a mean molecular pressure (over Boltzmann's
constant) for the clouds of 10$^{5}$ cm$^{-3}$\,K. Comparing this to the ionised boundary layer
pressures in Table \ref{tbl:ibl_results} reveals that the majority of the BRCs are potentially
underpressured with respect to their ionised exteriors. However, as mentioned in the previous
paragraph, it is possible that the mean molecular density of the clouds may be underestimated
due to optical depth effects or the fact that the $^{13}$CO ground-state transition
preferentially samples low-density gas. It is also possible that the $^{13}$CO at the molecular
boundary of the clouds   may be depleted by selective photodissociation (van Dishoeck \& Black
\cite{vdhb88}), leading to an underestimate of the true H$_{2}$ number density of the cloud. The
$^{13}$CO-derived densities have been shown to be almost an order of magnitude lower than the
sub-mm continuum-derived densities for three Northern Hemisphere BRCs (Thompson et
al.~\cite{thompson03a}), although the underlying cause of the depleted $^{13}$CO  abundance is
still unclear.

Taking all of these uncertainties into consideration we estimate that the mean
molecular density (and hence pressure) is likely to be underestimated by no more
than a factor of 15. Yamaguchi et al.~(\cite{ysmmmof99}) conclude that in the
worst-case scenario  their column density values are within 15\% of the true
value and in order to account for any possible selective photodissociation of
$^{13}$CO we  assume a  $^{13}$CO depletion factor of 10 (Thompson et
al.~\cite{thompson03a}). Using this scaling factor we estimate that the mean
molecular pressure is in reality closer to $\sim 1.5 \times 10^{6}$
cm$^{-3}$\,K. This molecular pressure is close to the ionised boundary layer for
the majority of our sample and we thus conclude that it is likely that most of
the clouds in Table \ref{tbl:ibl_results} are in pressure equilibrium. 

If the molecular pressure of the clouds is around $1.5 \times 10^{6}$
cm$^{-3}$\,K and most of the clouds are in pressure equilibrium,  then those 
clouds with the highest and lowest ionised boundary layer pressures are the
likeliest examples of clouds in pressure imbalance. SFO 58b and SFO 61 possess
IBL pressures of only $\sim 2 \times 10^{5}$ cm$^{-3}$\,K and are thus likely
to be underpressured with respect to their IBLs unless they are quiescent
(i.e.~low turbulent velocity and small observed FWHM linewidth) or of extremely
low density ($< 10^{3}$ cm$^{-3}$). These two clouds are thus in the initial 
evolutionary phase described by Lefloch \& Lazareff (\cite{ll94}) where the
photoionisation-induced shocks are stalled at the cloud surface until the
ionised boundary layer pressure equilibrates with the internal molecular cloud
pressure. Conversely SFO 75, SFO 77 and SFO 78 all possess IBL pressures in
excess of 10$^{7}$ cm$^{-3}$ and unless the molecular gas in these clouds has a
high velocity dispersion or density, it is unlikely that they are overpressured
with respect to their IBLs. These three clouds are the strongest candidates to
possess photoionisation-induced shocks propagating into their interiors.

We must stress however, that these results are based upon a global average value
for the mean cloud density derived from large-beam $^{13}$CO observations of
only 8 clouds. Higher resolution molecular line observations of individual
clouds in different molecular species are required to determine their molecular
densities, explore the isotopomer-selective photodissociation reported by van
Dishoeck \& Black (\cite{vdhb88}) and  investigate their pressure balance. We
have begun a study of this type with the Mopra 22m and SEST 15m millimetre-wave
telescopes and will report the results in forthcoming papers (Urquhart et
al.~\cite{urquhart2003}; Thompson et al.~\cite{thompson03b}).

\subsection{The candidate ionising stars of SFO 72 and 86}
\label{sect:sfo72_86}

As mentioned in Sect.~\ref{sect:uplim} the two clouds SFO 72 and SFO 86 possess predicted
ionising fluxes that are largely inconsistent with our observations. It is unlikely that these
inconsistencies can be explained by geometric effects alone;  for SFO 86 in particular the
star-cloud vector must be inclined by less than 10\degr\ to the line-of-sight to bring the
predicted ionising flux into line with the observational limit. 

We investigated wider-field optical and MSX 8 $\mu$m images of these two
sources to search for signs of any optical  extinction or mid-infrared emission
from hot dust in the HII region (which may absorb a significant fraction of the UV radiation
emitted from the stars) and to attempt to ascertain whether the
ionising star could be in a foreground or background position. If the star is
located in the foreground position then we would expect to see the bright
optical front face of the cloud perhaps combined with broad PDR 8$\mu$m
emission across the face of the cloud. If the star however lies behind the
cloud then we would expect the Earth-facing side of the cloud to be dark in the
optical but should still see a broad PDR positionally coincident with the
cloud as the cloud is likely to be optically thin to the  8$\mu$m emission . A
good example of this is SFO 68, where the optical rim of the cloud is partially
shielded by the dark molecular material, but the 8$\mu$m emission indicates
that there is a PDR located on the rearward face of the cloud.

\begin{center}
\begin{figure}
\includegraphics*[scale=0.5,trim=225 60 250 80]{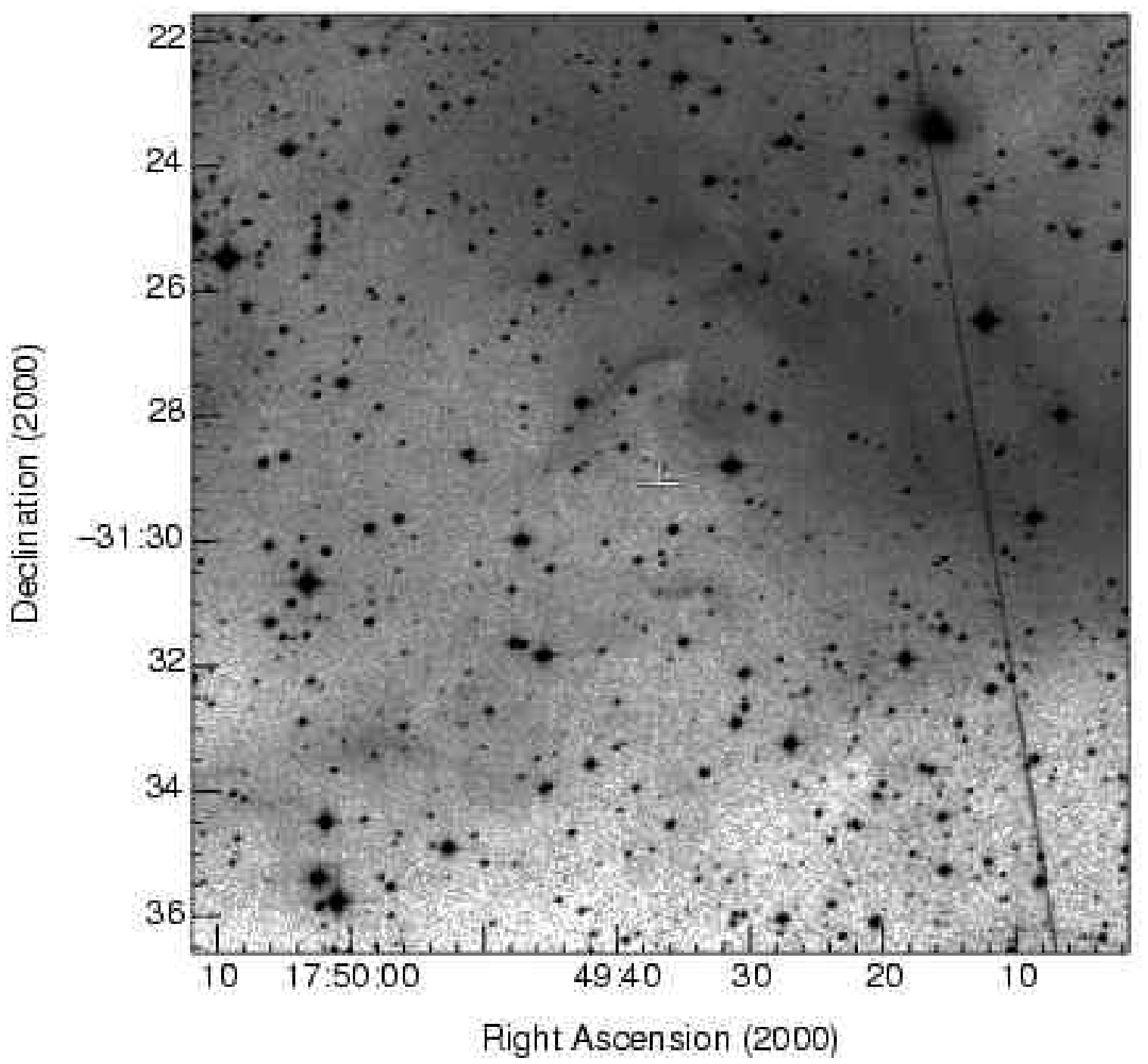}
\caption{Digitised Sky Survey red image of SFO 86. The greyscale coding is a
negative logarithmic scale to adequately present both the faint emission from the cloud and
the bright optical emission to the NW of the BRC. The IRAS point source
identified by Sugitani \& Ogura (\cite{so94}) is marked by a cross.}
\label{fig:sfo86dss}
\end{figure}
\end{center}

There is no detectable 8$\mu$m emission associated with SFO 86 or its surrounding HII region,
although there is a considerable amount of confusing emission in the red Digitised Sky Survey
image (see Fig.~\ref{fig:sfo86dss}). The cloud appears to be embedded in the parent HII region
and the  visible face of the cloud is optically bright, possibly indicating that the ionising
star is located in the foreground. It is impossible to ascertain the degree of inclination from
the optical image, but if the star from Table \ref{tbl:ionising_stars} is responsible for the
bright-rimmed appearance and ionisation of the cloud then it must be located at least 5 times
further from the cloud than its projected distance suggests. It is likely that BRC 86 is being
ionised at a much lower level than expected ($<$10$^{7}$ cm$^{-2}$\,s$^{-1}$) because the nearby
massive stars exciting the HII region are much more distant that previously suspected. 

\begin{figure}[h]
\includegraphics*[scale=0.75,trim=90 20 120 50]{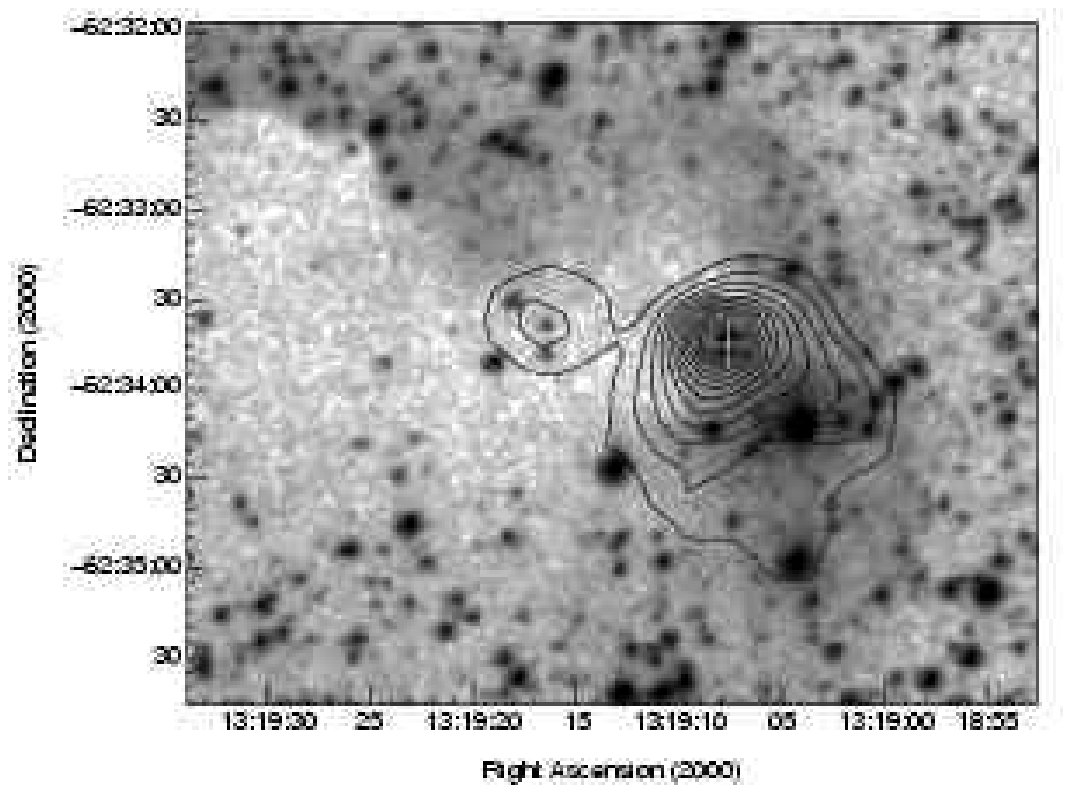}
\caption{Digitised Sky Survey red image of SFO 72 overlaid with contours of MSX
8 $\mu$m emission. The greyscale coding is a negative logarithmic scale to
better illustrate the high extinction seen towards the cloud and also the lack
of bright-rim emission from its edge. The cross indicates the position of SFO 72
from Sugitani \& Ogura (\cite{so94}). The MSX 8 $\mu$m emission is localised
around a potential PDR or cluster at the cloud tip.} 
\label{fig:sfo72dssmsxA}
\end{figure}

SFO 72 has a markedly different appearance to SFO 86 in the optical and 
mid-infrared (Fig.~\ref{fig:sfo72dssmsxA}). The cloud is the tip of a long
finger of extinction stretching for some 8\arcmin\ to the east before merging
into a larger area of extinction. A bright cluster of stars is visible at the
tip of the finger of extinction. The MSX 8$\mu$m emission is localised around
this cluster, signifying a compact PDR or hot dust emission. However, no radio
emission is associated with the finger or the star cluster down to a level of
10.8 mJy. The suspected ionising star of this cloud is HD 115455, an O7.5 III
star located 5\arcmin\ to the NW of the finger tip (and off the visible portion
of Fig.~\ref{fig:sfo72dssmsxA}).

We can draw several conclusions from the appearance of SFO 72 in the optical and
mid-infrared and its non-detection in the radio. The compact appearance of the
8$\mu$m emission suggests that it is centred upon either a compact HII region or
a PDR but the morphology of the 8$\mu$m emission is curved \emph{away} from the
finger-shaped cloud traced by the optical extinction, suggesting that the
emission is breaking free from the tip of the finger. The 8$\mu$m emission is
centred on the star CPD$-$61 3587, which is a member of the open cluster Stock
16 (Turner \cite{turner85}) and is variously reported as either an 07.5 III
(Lynga \cite{lynga69}) or a B2 V type star (Herbst \cite{herbst75}). The upper
limit of 10.8 mJy that we measured at 20 cm for this BRC rules out the
possibility that CPD$-$61 3587 is an O7.5 star, as from our upper flux limit 
Eq.~(\ref{eqn:ni}) suggests that we would have detected free-free emission from
an HII region excited by stars  earlier than type B1.

The edge of the finger-shaped cloud is not traced in either 8$\mu$m emission or
radio free-free emission, which suggests that the edges of this cloud are weakly
ionised at best ($\Phi \le 9 \times 10^{7}$ cm$^{-2}$\,s$^{-1}$). The edges of
the finger do not display strong bright-rimmed emission in the red Digitised Sky
Survey image and the overall appearance is that the cloud is illuminated on its
rearward face. SFO 72 is thus likely to lie in the foreground of its exciting 
star and if the responsible star is the O7.5 III giant  HD 115455 (Yamaguchi et
al.~\cite{ysmmmof99}) then the distance between the cloud and star must be
greater than 93.5 pc. Given this large distance and the fact that the 
E-W orientation of the finger does not align with the direction toward HD 115455
we suggest that it is not this star that has shaped the evolution of the finger
of SFO 72.

The 09.5 V star HD 115071 (also known as V961 Cen, LS 2998 or HIP 6437) is also
a member of the open cluster Stock 16 (Penny et al.~\cite{pgwsl02}) and is
located 21\arcmin\ to the west of SFO 72, which corresponds to a projected
distance of 11.6 pc. The ionising flux from this star at this assumed distance 
($\sim 7.5 \times 10^{7}$ cm$^{-2}$\,s$^{-1}$) is consistent with our radio
limits for that illuminating the cloud ($\le 9 \times 10^{7}$
cm$^{-2}$\,s$^{-1}$). The overall picture of this region is that the finger
assocated with SFO 72 lies some 100 pc in the foreground of the most massive
star powering the expansion of the HII region RCW 75, the O7.5 giant HD 115455.
The nearby O9.5 V star HD 115071, which is located some 11.6 pc to the
west, has been primarily responsible for shaping the morphology of the
finger-shaped cloud and possibly clearing the molecular material away from the
recently formed PDR surrounding CPD$-$61 3587 at the tip of the finger.

The description of the region surrounding SFO 72 underlines the powerful
technique of combining the optical, mid-infrared and radio images of these
regions in order to identify the stars responsible for ionising the BRCs, to
constrain their orientation and location with respect to the clouds and to
crucially measure the ionising radiation impinging upon the surface of the clouds. The
molecular material of the clouds is optically thin to the mid-infrared and radio
emission which means that the significant extinction and confusion present in
the optical images can be avoided. In addition the cross-correlation of the
radio and mid-infrared images allows those radio sources that are not associated
with rims of the BRCs (the Type 4 sources) to be removed from the sample. Future
studies aimed at the identification of bright-rimmed clouds should take note of
the advantages of the multi-wavelength approach in identifying the location and
orientation of these clouds to their exciting sources.

\subsection{The nature of the embedded sources}

In Sect.~\ref{sect:embedded} we showed that the far-infrared and radio
luminosities of the Type 3 and Type 2 radio sources identified in the survey are
consistent with the hypothesis that they are embedded massive YSOs or young 
stellar clusters (see Table \ref{tbl:type3}). Here, we will examine this
hypothesis in more detail so that we may explore the nature of the embedded
sources and investigate whether their formation can be linked to the action of
the nearby OB stars responsible for ionising the clouds.

\begin{figure}
\includegraphics*[scale=0.5,trim=240 70 240 90]{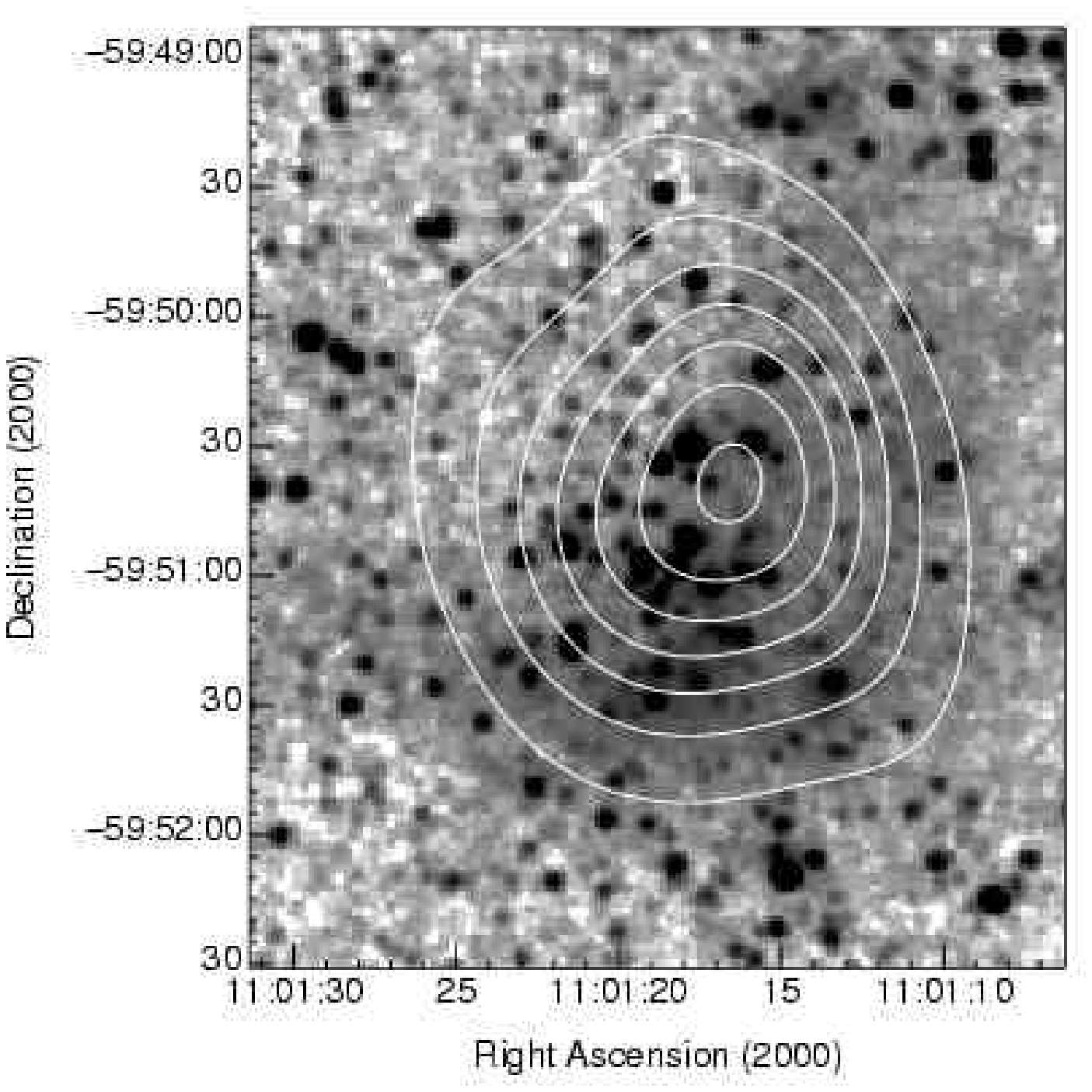}
\caption{2MASS $K_{s}$ band image of SFO 62, overlaid with the 3cm contours from
our radio survey. The contour levels begin at  6 mJy and are spaced by 10 mJy.}
\label{fig:sfo62_2mass}
\end{figure}

We carried out a search of the literature and the SIMBAD database of astronomical catalogues in order
to determine whether the embedded sources were previously discovered prior to our survey. We found that
almost all of the embedded sources (including the source SFO 82b, which was originally classified as a
Type 1 source) have been identified as embedded stellar clusters in a recent near-infrared survey of
the Milky Way (Bica et al.~\cite{bdb03}; Bica et al.~\cite{bdsb03}; Dutra et al.~\cite{dbsb03}). The
only source not identified in these catalogues is the broken-rimmed cloud SFO 62.  We present a 2MASS
$K_{s}$ band image of SFO 62 in Fig.~\ref{fig:sfo62_2mass}, which shows that the cluster of stars
located in the broken-rim  of the cloud  displays the same characteristics as the loose infrared
clusters identified by Dutra et al.~(\cite{dbsb03}). It is not known why SFO 62 does not appear in the
cluster catalogue of Dutra et al.~(\cite{dbsb03}), although SFO 62 is displaced by $\sim 6$\arcmin\
from the center of its 2MASS ``plate'' and perhaps for this reason was missed from the
5\arcmin$\times$5\arcmin\ fields that Dutra et al.~(\cite{dbsb03}) centred on each of the nebulae in
their search.

The literature search also revealed that SFO 59 and SFO 79 are respectively
associated with the infrared stellar clusters vdBH RN26 IRC and RCW 108 IRC
(Bica et al.~2003). SFO 74 is associated with two embedded infrared
sources LLN93 IRS 30-1 and 30-2, which are classified as having a spectral type
as early as B5 V (Liseau et al.~\cite{llnsm92}). As the IRAS sources associated
with these clouds (Sugitani \& Ogura \cite{so94}) follow the Wood \& Churchwell
(\cite{wc89}) IRAS colour  criteria for embedded ultracompact HII regions,
several of the embedded sources have been part of ultracompact HII region
surveys. SFO 59 and SFO 79 were surveyed for 6.7 GHz CH$_{3}$OH maser emission
by Walsh et al.~(\cite{whrb97}) with negative results. SFO 79 also formed part
of the search list of Bronfman et al.~(\cite{bnm96}) and was detected in
the J=2--1 line of CS, implying that the molecular density toward this region is
of the order 10$^{4}$--10$^{5}$ cm$^{-3}$.

All the embedded radio sources detected in our survey are thus revealed as
luminous ($L_{IR} \ge 4400$ L$_{\odot}$)  infrared stellar clusters that have
begun to ionise their surroundings and form compact HII regions. The clusters
must be relatively young as their IR and radio emission indicate that they
contain late O-type to early B-type stars, which have a main sequence lifetime
of $\sim 1.7 \times 10^{6}$ yrs. Most of the embedded regions are at best
marginally resolved in our radio survey (with the exception of SFO 59) and so we
cannot estimate the age of the compact HII regions by directly measuring their
radio diameter and estimating their expansion rate. However, it is unlikely that
they are in the ultracompact HII region phase because they have begun to clear
their surroundings sufficiently as to be visible in the near-infrared and
optical wavelength ranges, which we would not expect to be the case for the
deeply embedded ultracompact HII region phase (e.g.~Testi et al.~\cite{tfpr94}).
SFO 59 is extended on the scale of our $\sim 1$\arcmin\ FWHM beam and so this
object is certainly not an ultracompact HII region. The ultracompact phase lasts
for $\sim 10^{5}$ years and so we estimate that the age of the embedded clusters
is between 10$^{5}$ and $1.7 \times 10^{6}$ years. Higher resolution radio
observations are required to resolve the radio emission and determine the spatial
size of the embedded compact HII regions.

This maximum age is very similar to the age of the large HII regions (typically
1--2 Myr) at the edges of which the BRCs are found and so it is almost
impossible to determine whether the stars that are powering the HII regions are
responsible for inducing the formation of the stellar clusters. Hoever, even for
the more embedded clusters there has been sufficient time since the formation of
the massive stars powering the HII region for photoionisation-induced shocks to
travel deep inside the clouds and perhaps affect the star formation process. For
example, the cluster associated with SFO 79 is located approximately 3\arcmin\
from the bright rim of the cloud, which corresponds to a distance of 1.2 pc at
the assumed distance to SFO 79 (see Table \ref{tbl:ionising_stars}). The typical
shock velocities induced by the photoionisation of the BRC surface layers range
from 1--2 km\,s$^{-1}$ (Thompson et al.~\cite{thompson03a}; White et
al.~\cite{whiteeagle}) and at this velocity the shocks would require only $\sim
6 \times 10^{5}$ years to traverse the distance from rim to cluster.

Whilst we cannot conclude whether the formation of the embedded clusters was
due to the UV illumination from the nearby OB stars, we have nevertheless 
succeeded in showing that these BRCs are intimately involved in the star
formation process. At least 7 clouds from our original sample of 45 have been
shown to be associated with young stellar clusters containing high to
intermediate-mass stars (including SFO 72 and SFO 82b, which were originally
classed as a non-detection and a Type 1 rim source respectively). This is in
line with the higher IRAS luminosities found for BRCs over isolated
star-forming regions such as Bok globules (Sugitani et al.~\cite{smmto00},
\cite{sfo}), which suggest that BRCs may preferentially form stellar clusters
or significantly higher-mass stars than Bok globules.  Further work is required
in order to determine whether the formation of stars in these regions was
induced by the external action of the nearby OB stars, particularly
higher-resolution radio continuum observation to measure the spatial diameters
of the compact HII regions and constrain their ages and  optical/infrared
spectroscopy to provide more accurate age estimates for the stars contained in
the clusters. 

A more detailed theoretical treatment of the star formation processes in these
clouds is also needed, as the RDI models of Bertoldi (\cite{bertoldi}) and
Lefloch \& Lazareff (\cite{ll94}, \cite{ll95}) do not specifically address the
issue of star formation within these clouds. High-resolution molecular line and
millimetre/sub-millimetre continuum observations of the clouds could provide the
necessary physical parameters to model the star formation within these clouds
and potentially identify star-forming regions that were induced by the external
action of a nearby OB star.

\section{Summary and conclusions} 
\label{sect:concln}

We have carried out a radio-wavelength imaging survey of 45 bright-rimmed clouds
(BRCs) with the Australia Telescope Compact Array, with the aim of measuring the
physical conditions in their ionised boundary layers. We detected radio emission
from a total of 25 clouds and using a combination of optical Digitised Sky
Survey images and mid-infrared MSX 8$\mu$m images classified the radio emission
into 4 types. The four types are: Type 1 bright-rimmed clouds, Type 2
broken-rimmed clouds, Type 3 embedded radio sources and Type 4 unassociated
radio sources.

Type 1 sources are those where the radio emission follows the bright optical
rim of the cloud and correlates closely with the MSX 8$\mu$m emission, which is
a tracer of the photon-dominated surface layers of these clouds. Eighteen Type
1 clouds were detected in the survey and we evaluated the incident ionising
photon flux, electron density and ionised gas pressure for their ionised
boundary layers. With this sample we have more than doubled the existing number
of BRCs with known ionised boundary layer conditions (e.g.~Thompson et
al.~\cite{thompson03a}; White et al.~\cite{whiteeagle}; Megeath \& Wilson
\cite{mw97}; Lefloch et al.~\cite{llc97}).
Type 2 sources are those clouds where the radio and 8$\mu$m emission is
positionally correlated, but the optical appearance of the cloud resembles the
well known broken-rimmed globule CG 4 (Reipurth \cite{reipurth83};
Gonz\'alez-Alfonso et al.~\cite{g-acr95}). Only one Type 2
source was identified (SFO 62).  

Type 3 sources are strong radio emitters that are both positionally coincident
with luminous mid-infrared sources and embedded within the bright-rimmed clouds
rather than located at the edge. The high infrared and radio luminosities of
these objects imply that they are embedded stellar clusters containing late O
and early B-type stars. Four type 3 sources were detected in the survey. Type 4 sources are radio sources that are either
not positionally coincident with the optical or mid-infrared emission tracing
the clouds or whose morphology does not follow the optical or mid-infrared
morphology of the clouds. It is possible that a number of Type 4 sources represent
free-free emission associated with the clouds but observations of 
higher sensitivity and resolution than our survey are required to make a
definitive statement.

We draw the following conclusions from our survey:

\begin{enumerate}

\item Comparing the pressure of the ionised gas in the ionised boundary layer to
a global estimate of the  internal molecular pressure of  BRCs indicates that the
majority of the clouds are  in pressure equilibrium. It is thus highly likely
that photoionisation-induced shocks are currently propagating into these clouds,
in line with the predictions of the radiative-driven implosion (RDI) models of
Lefloch \& Lazareff (\cite{ll94}) and Bertoldi (\cite{bertoldi}).

\item The two clouds SFO 58b and SFO 61a have the lowest ionised boundary layer
pressures and are thus the most probable clouds where the interior  molecular
pressure could exceed the exterior ionised gas pressure. These clouds are
presumably in the evolutionary phase prior to that of pressure balance between
their interior and exterior (Lefloch \& Lazareff \cite{ll94}). In this phase the
photoionisation-induced shocks are stalled at the ionised boundary of the cloud
until the pressure in the ionised boundary layer rises and equilibrates with the
interior molecular pressure of the cloud.

\item In most cases the physical properties of the ionised boundary layers are
consistent with their expected ionisation from the OB stars responsible for
exciting the large HII regions at the edges of which the clouds are found. Where
the predicted ionising flux illuminating the clouds is greater than that derived
from the observations we have shown that differences of a factor of a few can be
readily explained by projection effects. For a number of clouds the predictions
and observations cannot be reconciled and we suggest that there are additional
ionising sources present within the HII regions. A SIMBAD database search
revealed several additional OB stars but their overall contribution to the
ionising flux is still insufficient to match the observations. Either the
spectral types of the known stars are earlier than estimated or there are
additional ionising stars located within the HII regions.

\item Using a multi-wavelength approach combining the radio, mid-infrared and
optical data, we show that the ionising star of SFO 72 is not the O7.5 star HD
115455,  as originally thought, but is more likely to be the O9.5 star HD
115071, which lies some 21\arcmin\ to the west of SFO 72. The non-detection of this
cloud in the radio suggests that it is located at least 93.5 pc in the
foreground of HD 115455. An embedded cluster is currently breaking free from the
molecular gas at the tip of SFO 72 and from our radio upper limit we constrained
the earliest star within this cluster to be B1, consistent with the observations
of Herbst (\cite{herbst75}).

\item The 4 embedded Type 3 radio sources  detected in the survey are shown to
be embedded stellar clusters containing late O and early B-type stars. All of
the embedded sources are indentified with infrared stellar clusters found in
recent 2MASS surveys (Bica et al.~\cite{bdb03}; Bica et
al.~\cite{bdsb03}; Dutra et al.~\cite{dbsb03}) or in 2MASS images downloaded as
part of this paper (see Fig.~\ref{fig:sfo62_2mass}). Three more BRCs are also
found to be associated with stellar clusters which have either disrupted their
natal molecular cloud (SFO 62) or are on the verge of breaking free from the
molecular gas (SFO 72 \& SFO 82b). Each of these clusters is sufficiently 
luminous in the far-infrared or radio to contain at least early B-type stars and
thus we have shown that bright-rimmed clouds are implicated in the formation of
high to intermediate mass-stars. However, further observations and modelling are
required to investigate whether this star formation may have been induced by the
external action of the OB star or stars ionising the bright-rimmed clouds.
\end{enumerate}

\begin{acknowledgements} 
The authors would like to thank the Director and staff
of the Paul Wild Observatory, Narrabri, New South Wales, Australia for their 
hospitality and assistance during the observing run. We would also like to thank
Michael Burton for useful discussions regarding the mid-infrared emission from
the clouds.  MAT is supported by a PPARC postdoctoral grant and JSU by a PPARC
doctoral  studentship. This research would not have been possible without the
SIMBAD astronomical database service operated at CCDS, Strasbourg, France and
the  NASA Astrophysics Data System Bibliographic Services. The Digitized Sky
Survey was produced at the Space Telescope Science Institute under U.S.
Government grant NAG W-2166. The images of these surveys are based on
photographic data obtained using the Oschin Schmidt Telescope on Palomar
Mountain and the UK Schmidt Telescope. The plates were processed into the
present compressed digital form with the permission of these institutions.
Quicklook 2.2 $\mu$m 2MASS images were obtained as part of the Two Micron All
Sky Survey (2MASS), a joint project of the University of Massachusetts and the
Infrared Processing and Analysis Center/California Institute of Technology,
funded by the National Aeronautics and Space Administration and the National
Science Foundation. This research made use of data products from the Midcourse
Space  Experiment.  Processing of the data was funded by the Ballistic  Missile
Defense Organization with additional support from NASA  Office of Space
Science.  This research has also made use of the  NASA/ IPAC Infrared Science
Archive, which is operated by the  Jet Propulsion Laboratory, California
Institute of Technology,  under contract with the National Aeronautics and
Space  Administration. 
\end{acknowledgements}

\end{document}